\documentclass[letterpaper,twoside,sort&compress,super,3p,12pt]{elsarticle-BMJ}
\usepackage{fancyhdr}
\usepackage{graphicx}
\usepackage{amsmath,amssymb,mathtools}
\usepackage{epstopdf}
\usepackage[table]{xcolor} 
\usepackage{color}
\usepackage[flushleft]{threeparttable} 
\usepackage{multirow, fancybox}
\usepackage{booktabs}
\usepackage{pdflscape} 
\usepackage{paralist} 
\usepackage{url}
\usepackage{cleveref}
\renewcommand{\theequation}{\arabic{section}.\arabic{equation}}
\renewcommand{\thefigure}{\arabic{section}.\arabic{figure}}
\renewcommand{\thetable}{\arabic{section}.\arabic{table}}

\biboptions{comma}
\makeatletter
\renewcommand\@biblabel[1]{#1} 
\makeatother

\newif\ifnotesw \noteswtrue

\newcommand{\CHK}{Check It }
\newcommand{\CHKn}{Check It} 

\newcommand{\pt}{{\cdot}}
\begin{document}
\begin{frontmatter}
\title{Modelling the Impact of Screening Men for \textit{Chlamydia Trachomatis} on the Prevalence in Women}
\author[Tmath,UTSA]{Zhuolin Qu}
\ead{zhuolin.qu@utsa.edu}
\author[ASU]{Asma Azizi}
\ead{soodeh.azizi@gmail.com}
\author[Tph]{Norine Schmidt}
\ead{nschmid1@tulane.edu}
\author[Tph]{Megan Clare Craig-Kuhn}
\ead{mcraigkuhn@tulane.edu}
\author[Tecon]{Charles Stoecker}
\ead{cfstoecker@tulane.edu}
\author[Tmath]{James M Hyman}
\ead{mhyman@tulane.edu}
\author[Tph]{Patricia J Kissinger}
\ead{kissing@tulane.edu}
\address[Tmath]{Department of Mathematics, Tulane University, New Orleans, LA 70118}
\address[UTSA]{Department of Mathematics, The University of Texas at San Antonio, San Antonio, Texas, 8249}
\address[ASU]{Simon A. Levin Mathematical Computational Modeling Science Center, Arizona State University, Tempe, AZ, 85281}
\address[Tph]{Department of Epidemiology, Tulane University, New Orleans, LA }
\address[Tecon]{Department of Health Policy and Management, Tulane University, New Orleans, LA 70112}

\begin{abstract}
\paragraph{Background}
\textit{Chlamydia trachomatis} is the most commonly reported infectious disease in the United States and causes important reproductive morbidity in women. The Centers for Disease Control and Prevention have recommended routine screening of sexually active women under age 25 but have not recommended screening among men. Consequently, untested and untreated men may serve as a reservoir of infection in women. Despite three decades of screening women, the chlamydia prevalence has continued to increase. Moreover, chlamydia is five times more common in African American (AA) youth compared to Whites, constituting an important health disparity.

\paragraph{Methods}
The \CHK program is a bundled Ct intervention targeting AA men aged 15-24 who have sex with women. We created an individual-based network model to simulate a realistic chlamydia epidemic on sexual contact networks for the target population. Based on the practice in \CHKn, we quantified the impact of screening young AA men on the chlamydia prevalence in women. 

\paragraph{Findings}
We used sensitivity analysis to quantify the relative importance of each \CHK intervention component, and the significance ranked from high to low was venue-based screening, expedited index treatment, expedited partner treatment, rescreening. We estimated that by annually screening $7\pt5\%$ of the target male population, the chlamydia prevalence would be reduced by $8\pt1\%$ and $8\pt8\%$ in men and women, respectively.

\paragraph{Interpretation}
The findings suggested that male-screening has the potential to significantly reduce the prevalence among women.

\paragraph{Funding}
National Institutes of Health, National Institute of Child Health and Human Development, Tulane University.

\end{abstract}

\end{frontmatter} 
\thispagestyle{fancy}
\newpage
\section*{Research in context}
\subsection*{Evidence before this study} 
\textit{Chlamydia trachomatis} is the most commonly reported infectious disease in the US. Women more commonly experience severe sequelae, such as infertility, pelvic inflammatory disease, and ectopic pregnancy. For three decades, the Centers for Disease Control and Prevention (CDC) have recommended routine screening of sexually active women $<25$ years old, together with providing her and her partner(s) with treatment and rescreening, as part of preventive healthcare. However, the prevalence has continued to rise. In 2018, there were about $1\pt8$ million cases reported to the CDC, and the rate has increased by 19\% since 2014. Recommendations of routine screening don't exist for men in the US, who may serve as a reservoir of infection in women, and few efforts have been made to evaluate the potential impact of male-screening on prevalence in women. 

Most chlamydia simulation studies use differential equation-based compartmental models. These models cannot capture the complex assortative mixing in sexual partnership networks. The structure of the sexual network affects the spread of the infection and the effectiveness of the mitigation efforts. 

\subsection*{Added value of this study}
Our stochastic, heterosexual, and individual-based model provided a framework for a public health team to answer ``what if'' questions that are hard to address in the field. Our model simulated the spread of chlamydia through dynamic sexual networks that were embedded in a social contact network. The sexual networks were generated from survey data sets and captured the sexual behaviour among an assortative mixing population in New Orleans. We first simulated the current chlamydia epidemic under the baseline healthcare interventions, including routine screening in women and treatment for symptomatic cases in men and women during clinic visits. We then assessed the impact of a new male-screening bundled strategy, including expedited index treatment, expedited partner treatment, and rescreening, on the current chlamydia prevalence in women. To our knowledge, our study is the first modelling analysis to investigate this intervention strategy in this population cohort.

\subsection*{Implications of all the available evidence}
Our modelling study confirmed that the present screening activities for women alone cannot control the current chlamydia epidemic. The results suggested that adding the male-screening with the expedited index and partner treatment has the potential to significantly reduce the chlamydia prevalence among women. 
Our findings also show that expedited partner treatment becomes more effective in reducing the chlamydia prevalence in women as the coverage of male-screening increases.

\newpage
\section{Introduction}
\subsection{Chlamydia and current screening policy}
\textit{Chlamydia trachomatis} (Ct) is the most commonly reported infectious disease in the US, with about $1\pt8$ million cases each year \cite{CDC2019Sexually}. It is a major cause of infertility, pelvic inflammatory disease, and ectopic pregnancy among women \cite{hillis1996Screening} and has been associated with increased HIV acquisition \cite{ward2010contribution}. Because women experience the most severe sequelae, the focus of Ct prevention in the US has been on screening sexually active women $<$ 25 years old, providing her and her partner(s) with treatment and rescreening.

There is no recommendation for Ct screening among men in the US. In 2007, an expert panel at the Centers for Disease Control and Prevention (CDC) concluded that the evidence is not sufficient to recommend routine screening for Ct in sexually active young men \cite{report2007male}. The conclusion was based on the Ct prevalence in 2007 and the feasibility, efficacy, and cost-effectiveness of screening men. However, since then, evidence of the benefit of screening young men for Ct in high prevalence areas has been mounting. More recent studies show that screening men can be cost-effective \cite{gift2008program} and significantly reduce Ct prevalence in women \cite{gopalappa2013cost}. 

\subsection{\CHK program and intervention strategies}
The community-based program ``\CHKn'' \cite{KissingerA71} targets African American (AA) men aged 15-24 in New Orleans. The core of this intervention is Ct screening for men, and we hypothesise that men are an important reservoir of infection for women and need to be targeted for intervention \cite{KissingerA71}. 
The \CHK program bundles several key Ct prevention strategies:
\begin{itemize}
\item[-] Venue-based screening (VBS) by recruiting participants at non-clinical community venues, such as barbershops, colleges, and universities, in high-prevalence neighbourhoods characterised by similar demographic and geographic factors. This venue-based enrolment is enhanced with marketing strategies, such as the distribution of flyers, web education, social media, and informational cards.
\item[-] Expedited treatment by providing treatments for the Ct+ men (or index) (expedited index treatment or EIT) and his sexual partner(s) (EPT) via partnering community pharmacies without a medical examination to speed up treating his sexual partners and reduce reinfection rates in the index.
\item[-] Rescreening, where Ct+ men are retested for infection three months after treatment.
\item[-] Social network peer referral (SNPR) that encourages men to refer young AA men in their social network to \CHK via flyers, social media, or text messages to promote the program and increase the total enrolment.
\end{itemize}

\subsection{Mathematical modelling}
Mathematical models create frameworks for understanding the underlying epidemiology of disease and help test the potential effectiveness of different approaches to bring the epidemic under control. Our modelling effort created a detailed simulation to model the specific practices implemented by \CHK program. The model predicted the impact of \CHK on mitigating the Ct epidemic among AA women by screening young AA men at different intervention intensities.

\section{Methods}
We used a stochastic agent-based model to simulate the transmission of Ct among the 15-24-year-old AA New Orleans population and quantify the impact of the \CHK program on the Ct prevalence in women. 

We first generated dynamic heterosexual networks of a synthetic population, which were embedded in a grand social contact network (\cref{sec:generation}). We then modelled the Ct transmission pathways through these sexual networks (\cref{sec:SIS}). The model kept track of the infection status and sexual behaviour for each individual in the population. The generated dynamic sexual networks are based on the data from two survey studies \cite{KissingerA71,greenWP58} that investigated the sexual behaviour of the men and women targeted for the study.

We then modelled the intervention strategies, including both the standard preventive healthcare for women and the \CHK intervention (\cref{sec:intervention}), for each individual to study the impact of male-screening on the Ct prevalence in women. 

\subsection{Generation of a synthetic population over dynamic sexual networks \label{sec:generation}}
Similar to the approach in \citet{azizi2018generating}, we constructed a closed 5000-member synthetic population, where the heterosexual partnerships were represented by bipartite sexual networks. The heterosexual networks reflected the assortative mixing pattern among our targeted population by matching the population-level quantities from two surveys in New Orleans: the ongoing \CHK study \cite{KissingerA71} and the ``You Geaux Girl!'' (YGG) study \cite{greenWP58}, which enrolled 1318 AA men (as of April 2019, age-range 15-24) and 649 AA women (age range 18-19), respectively.

We categorised one's partner into either a primary partner or a casual partner. The distributions of primary and casual partnerships were based on the survey answers (detailed criteria in \cref{sec:A12}). These heterosexual partnerships can be asymmetric: A is B's primary partner, but B may be A's casual partner. Thus, this approach results in three types of partnerships: primary-primary, primary-casual, and casual-casual partnership. 

When modelling the epidemics over a long time, we updated the sexual partner(s) every two months (the time-frame covered by the \CHK survey). This created a series of dynamic sexual networks. To simulate a realistic partner updating behaviour, these sexual networks are embedded in a grand social contact network for 150,000 people in New Orleans generated by the Simfrastructure agent-based modelling and simulation system \cite{NDSSL2008synthetic}. During the updating process, the primary-primary partnerships were preserved throughout the simulations, and we updated one's casual sexual partner(s) through his/her social connections in the closed 5000 synthetic population. Specifically, half of the primary-casual and all the casual-casual partnerships are replaced by one's social contacts from the background social network every two months. Overall, we maintained the same 5000 population, and approximately 80\% of the partnerships were from one's heterosexual social network, in agreement with the \CHK data.

We give additional details on the generation of the dynamic sexual networks embedded in the social network in \cref{sec:A_network}. 

\subsection{Chlamydia epidemic on dynamic sexual networks \label{sec:SIS}}
We modelled the infection status of each individual using the Susceptible-Infectious-Susceptible (SIS) framework. All uninfected individuals are susceptible to being infected, and all infected people recover to this susceptible state after either spontaneous recovery or treatment. A susceptible individual can be infected by his/her infectious partner. The force of infection is the probability (per day) that a susceptible will be infected. This probability was estimated by considering risk factors of how many partners the person has, the type of partnership with each partner, the probability of having sexual contact per partner per day, and the probability of using a condom. 

We estimated the sexual contact rates for the primary and casual partnerships from the datasets (\cref{tab:contact_men_women}). We observed that the contact rate was higher for primary partnerships than for casual partnerships, and there was a decreasing trend in the per partner contact rate when the number of partners increases. Moreover, the datasets gave a higher probability of using a condom with a casual partner than with the primary one (\cref{tab:parameters}). We introduced a condom failure rate to include cases when the condom is not used properly. 

The details on the model configuration are fully described in \cref{sec:A_SIS}.

\subsection{Modelling the intervention strategies \label{sec:intervention}}
Our goal is to investigate the net impact of screening men through the \CHK program given the existing screening policy for women and the ongoing endemic Ct epidemic. To this end, we outlined the intervention strategies for both women and men in separate flow charts in \cref{fig:intervention}. We used solid lines to illustrate the \CHK intervention approaches and the dashed lines to represent the current healthcare interventions. The baseline scenario accounts for the Ct screenings completed at women's annual exams as part of regular preventive healthcare. We also included the Ct screenings prompted by symptomatic infections (clinical visits) in the baseline scenario for both men and women. 

\subsubsection{Existing intervention strategies}
The current (baseline) Ct mitigation efforts for women (the right side of \cref{fig:intervention}) include Ct screening during the routine annual exams and clinical visits for symptomatic infections. The model assumed that the same fraction ($\sigma_a^w$) of women return for a physical exam each year. Symptoms can appear in a small fraction of infected women ($\sigma_s^w$), and we assumed that a fraction $\theta_s$ of these women get medical care within an average of $\tau_s$ days, including incubation period and appointment scheduling process, after infection \cite{farley2003asymptomatic}. After the diagnostic test, we assumed that all the positive cases get index treatment in an average of $\tau_t^w$ days. Moreover, the CDC recommends doing EPT for the infected women by providing treatment to the patient to bring to her partner(s) without first examining the partner(s) \cite{CDCept}. We assumed that a fraction $\theta_p^w$ of the partners are treated with an average delay of $\tau_p^w$ days. The EPT fraction $\theta_p^w$ is the product of (1) the fraction of the physicians practising the EPT as recommended and (2) the fraction of compliance from the notified partners. Last, women diagnosed with the Ct infection should be retested after the initial treatment \cite{CDCtreat}. Thus, we assumed a fraction $\theta_r^w$ of treated women are retested for infection $\tau_r^w$ days after the initial treatment.

Since routine male-screening is not recommended and is rarely practiced \cite{st.lawrence2002STD}, we only account for screenings prompted from symptomatic infections. Similar to the process in women, we assumed that there are a small fraction ($\sigma_s^m$) of infected men that develop symptomatic Ct infections, and, with a delay of $\tau_s$ days, a fraction $\theta_s$ of them get screening and treatment. Moreover, follow-up interventions, such as partner treatment or rescreening after the index treatment, are hardly implemented in clinics for men \cite{st.lawrence2002STD}. Therefore, follow-up interventions for men were not included in the model.

\subsubsection{\CHK intervention strategies}
The \CHK program recruited participants in community venues, including community colleges, historically black colleges and universities, barbershops, and other community-based organisations. This venue-based enrolment is enhanced with the marketing strategies (distribution of flyers, web education, social media, and informational cards) and enrols a fraction $\sigma^m_e$ of the target male population for Ct screenings.

Among this venue-based enrolment, some participants learn about the program through their social networks, such as text messages and information cards sent by friends or word-of-mouth. We combined such peer impact as an SNPR and included it as a source of enrolment. On average, the proportion between the non-peer enrolled men and peer-recruited men is $1:\rho$, and we assumed that peer-referred men are enrolled in the program with an average delay of $\tau_n^m$ days.

We modelled the non-peer enrolment process as a random sampling from the entire male population. Meanwhile, we modelled the peer-referred enrolment (SNPR) by searching the background social network of each non-peer enrolled man and randomly sampling among the eligible candidates.

The rest of the intervention practice was modelled similarly to women's case: a fraction $\theta_t^m$ of the screened and infected men receive EIT after a delay of $\tau_t^m$ days. A fraction $\theta_p^m$ of these men's partners receive EPT with a delay of $\tau_p^m$ days. Finally, a fraction $\theta_r^m$ of the index men return for rescreening $\tau_r^m$ days after the initial infection. If the rescreened men are infected with Ct, then they are treated as index cases, and the intervention process is repeated. 

The \cref{sec:A_SIS} and \cref{sec:A_Female} provide more details on the modelling.

\subsection{Model calibration and parameters}
\begin{landscape}
\begin{figure}[htbp]
\centering
\includegraphics[width=1.2\textwidth]{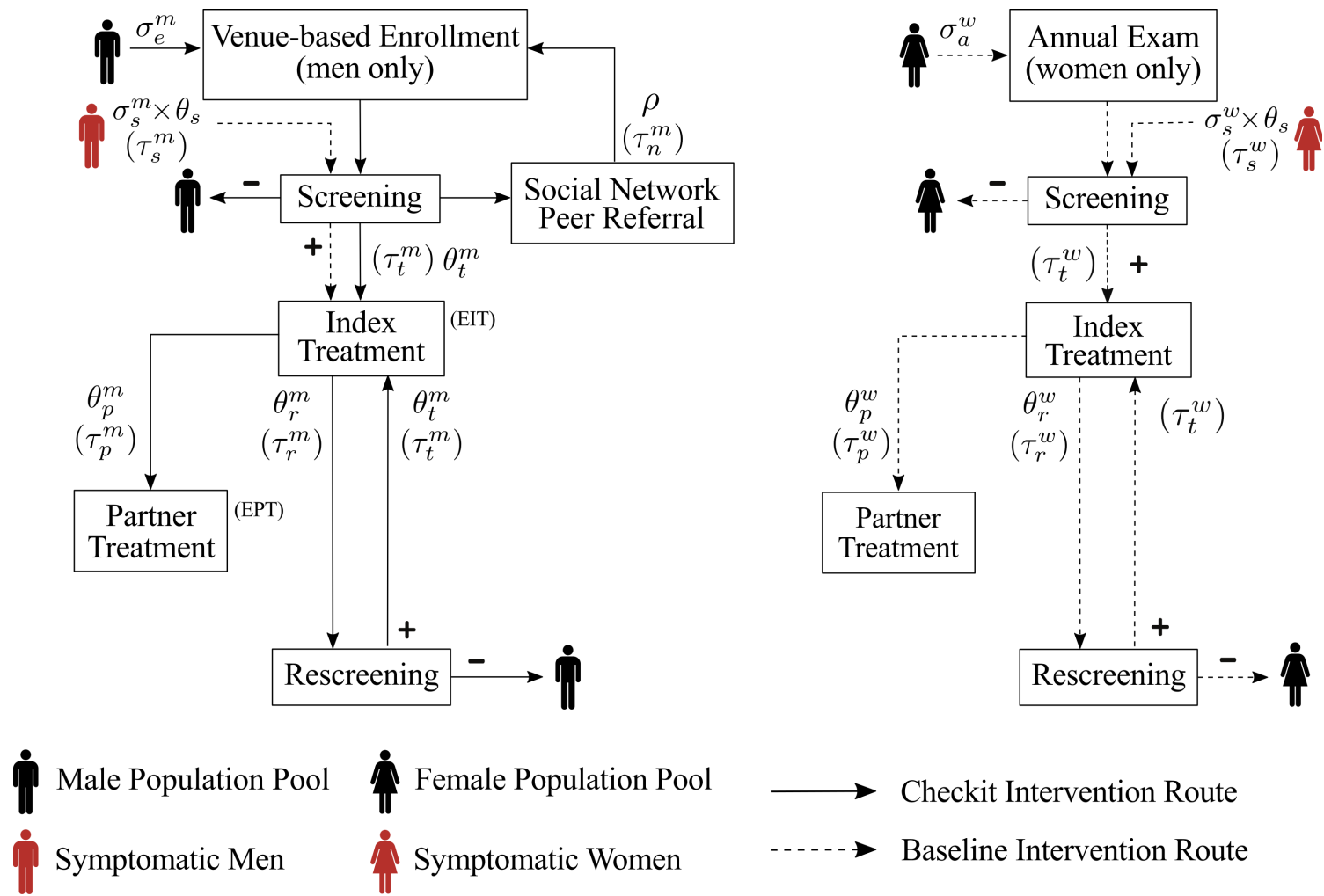}
\caption{Flowchart for Ct intervention strategies in men and women. The solid lines are the \CHK intervention approaches, and the dashed lines are the existing interventions implemented in the healthcare system before the \CHK program. These current interventions include women's annual screening and screenings prompted by symptomatic infections in both men and women. Our modelling effort examined the net impact of adding the \CHK program to help control the Ct epidemic. The \CHK program targets the male population and uses venue-based enrolment, expedited index treatment (EIT), expedited partner treatment (EPT), rescreening, and social network peer referral (SNPR).\label{fig:intervention}}
\end{figure}
\end{landscape}

We distributed the initial infections in the population consistent with the distribution of an emerging epidemic. We obtained this quasi-steady-state balanced initial condition by starting a small epidemic in the past and letting it grow to the current (pre-\CHKn) Ct endemic state (see \cref{sec:A_IC}). This initialization process considered the existing (baseline) combination of Ct interventions (dashed routes in \cref{fig:intervention}) to give a comprehensive approximation of the current Ct control before the launch of \CHKn. 

The model parameters used in the simulations (\cref{tab:parameters}) represent our best knowledge of the current epidemic. The parameters for the Ct transmission probabilities per sexual contact between men and women ($\beta^{m2w}$ and $\beta^{w2m}$) are only known with large uncertainty. We used these as tuning parameters to calibrate the model so that the initial balanced condition matched the current Ct prevalence in New Orleans among the target population ($10\pt2\%$ in men and $13\pt5\%$ in women among 15-24-year-old AA young adults \cite{torrone2014prevalence}).

Approximately half of the Ct infections are cleared naturally by the first year after being infected and $80\%$ are cleared after two years. We fitted an exponential distribution for the average time of natural recovery $\tau_n$ days to the data from \citet{molano2005natural}. We assumed the recovery time with treatment also follows an exponential distribution $\tau_t \sim exp(1/7)$ in days.

\subsection{Sensitivity analysis}
The model parameters in \cref{tab:parameters} represent the best-guess estimates for practical scenarios, and we used sensitivity analysis to quantify the most significant model parameters \cite{chitnis2008determining}. To check the impact of each component of the \CHK intervention bundle one-at-a-time, we varied each intervention parameter (the parameter of interest or POI), while fixing the other parameters. We then checked its corresponding impact on the Ct prevalence for women, men, and the overall population (the quantities of interest or QOIs). 

We defined the normalised relative sensitivity index (SI) of a QOI, $q(p)$, with respect to the POI, p, as $\mathcal{S}_p^q$ = $p/q \times \partial q/\partial p$. This SI, $\mathcal{S}_p^q$, measures the percentage change in the QOI given the percentage change in an input POI. If parameter $p$ changes by $x\%$, then quantity $q$ changes by $\mathcal{S}_p^q\times x\%$.

To further investigate the synergistic effect of the intervention components beyond the current levels, which could be limited by the protocols and available resources of \CHKn, we conducted the global sensitivity analysis by varying two intervention parameters together while fixing all other parameters. We then predicted the impact under different combinations of intervention parameters. 

\subsection{Role of the funding source}
The funders had no role in study design, data collection, data analysis, data interpretation, or writing of the report. The corresponding author had full access to all the data in the study and had final responsibility for the decision to submit for publication.

\begin{table}[ht!]
\hspace{-1.3cm}
\begin{threeparttable}
\centering
\begin{tabular}{@{}m{0.4cm} m{10cm} m{0.7cm} m{1.85cm} m{4.1cm}@{}}
\toprule
 & Description & Base. & Range & Reference\\
\midrule
${\beta^{{\scriptscriptstyle m2w}}}$ & ~Transmissibility from men to women per contact & $0\pt30$ & $0\pt04$ - $0\pt5$\tnote{*} & \citet{kretzschmar1996modeling},\newline \citet{turner2006developing}\\
$\beta^{{\scriptscriptstyle w2m}}$ & ~Transmissibility from women to men per contact & $0\pt10$ & $0\pt04$ - $0\pt25$\tnote{*} & \citet{kretzschmar1996modeling},\newline \citet{turner2006developing}\\
$\tau_n $ & Average time to recovery without treatment (years) & $1\pt32$ & exponential &\citet{molano2005natural} \\
$\tau_t $ & Average time to recovery after treatment (days) & 7 & exponential & \citet{martin1992controlled} \\
$c_p$ & Fraction of condom use for primary partners & $0\pt54$ & $\cdot\cdot$ &\CHK\cite{KissingerA71} \\
$c_c$ & Fraction of condom use for casual partners & $0\pt66$ & $\cdot\cdot$ & \CHK\cite{KissingerA71}\\
$c_\epsilon$ & Condom failure rate & $0\pt1$ & $\cdot\cdot$ &\citet{trussell2007choosing}\\
$\theta_s$ & Fraction of symptomatic infections screened & $0\pt7$ & $0\pt6$ - $0\pt8$ & \citet{farley2003asymptomatic}\\
$\tau_s$ & Time lag in screening for symptomatic infection (days) & 21 & 14 - 28 &CDC\cite{CDCdetail}\\
\midrule
 & Women's intervention parameters& & & \\
\midrule
$\sigma^w_a$ & Fraction of the target women who are screened annually & $0\pt6$ & $0\pt56$ - $0\pt65$ & \citet{wiehe2011chlamydia},\newline \citet{hoover2014chlamydia}  \\
$\sigma^w_s$ & Fraction of symptomatic infection in women & $0\pt3$ & $\cdot\cdot$  &\citet{farley2003asymptomatic} \\
$\theta_p^w $ & Fraction of partner treatment for index women & $0\pt24$ & $\cdot\cdot$	& Derived \\
&- Fraction of physicians practicing partner treatment & $0\pt4$ & $0\pt3$ - $0\pt5$ & \citet{hogben2005patient}\\
&- Fraction of compliance for partner treatment & $0\pt6$ & $0\pt4$ - $0\pt8$ & \citet{golden2005effect},\newline \citet{schillinger2003patient}\\
$\theta_r^w $ & Fraction of treated women who are rescreened & $0\pt2$ & $0\pt17$ - $0\pt28$ &\citet{xu2011use}\\
$\tau_t^w$ & Time lag in treatment for screened women (days)   & 2 & $\cdot\cdot$ & \citet{Cttest}\\
$\tau_p^w$ & Time lag in partner treatment for treated women (days) & 6 & 0 - 15 & \citet{golden2005effect}\\
$\tau_r^w$ & Time lag in rescreening for treated women (days) & 105 & 80 - 130 & CDC\cite{CDCtreat}, \citet{xu2011use}\\
\midrule
& Men's intervention parameters& & & \\
\midrule
$\sigma^m_e$ & Fraction of target population enrolled per year & $0\pt075$ & $\cdot\cdot$ & \CHK\cite{KissingerA71} \\
& - Fraction of non-peer VBS-enrolment & $0\pt76$ & $\cdot\cdot$ & \CHK\cite{KissingerA71} \\
& - Fraction of SNPR enrolment & $0\pt24$ & $\cdot\cdot$ & \CHK\cite{KissingerA71} \\
$\rho$ & Number of peer-recruited men per VBS-enrolled man & $0\pt32$ &$\cdot\cdot$ & Derived\\
$\sigma^m_s$ & Fraction of symptomatic Ct infection in men & $0\pt11$ & $\cdot\cdot$ & \citet{farley2003asymptomatic}\\
$\theta_t^m $ & Fraction of screened positive men treated (EIT) & $0\pt76$ & $0\pt1$ - $0\pt9$  & \CHK\cite{KissingerA71} \\
$\theta_p^m $ & Fraction of partner treatment for index men (EPT) & $0\pt27$ & $0\pt1$ - $0\pt9$ & \CHK\cite{KissingerA71}\\
$\theta_r^m $ & Fraction of treated men with rescreening (\%) & $0\pt08$ & $0\pt1$ - $0\pt9$ & \CHK\cite{KissingerA71} \\
$\tau_t^m$ & Time lag in treatment for screened men (days)  & 12 & $\cdot\cdot$ & \CHK\cite{KissingerA71}\\
$\tau_n^m$ & Time lag in screening for men enrolled via SNPR (days) & 7 & $\cdot\cdot$  &Assume\\
$\tau_p^m$ & Time lag in partner treatment for treated men (days) & 2 & $\cdot\cdot$ & \CHK\cite{KissingerA71}\\
$\tau_r^m$ & Time lag in rescreening for treated men (days) & 102 & $\cdot\cdot$ & \CHK\cite{KissingerA71}\\
\bottomrule
\end{tabular}
\caption{Model parameters.\label{tab:parameters}}
\begin{tablenotes}
\item[*] Values come from mathematical modelling papers where the transmission rates were estimated by least square fitting to the corresponding Ct prevalence data ($10\pt2$\% in men, $13\pt5$\% in women).
\end{tablenotes}
\end{threeparttable}
\end{table}

\section{Results}
\subsection{Impact of \CHK at existing intervention level}
\begin{figure}[htbp]
\centering
\includegraphics[width=0.6\textwidth]{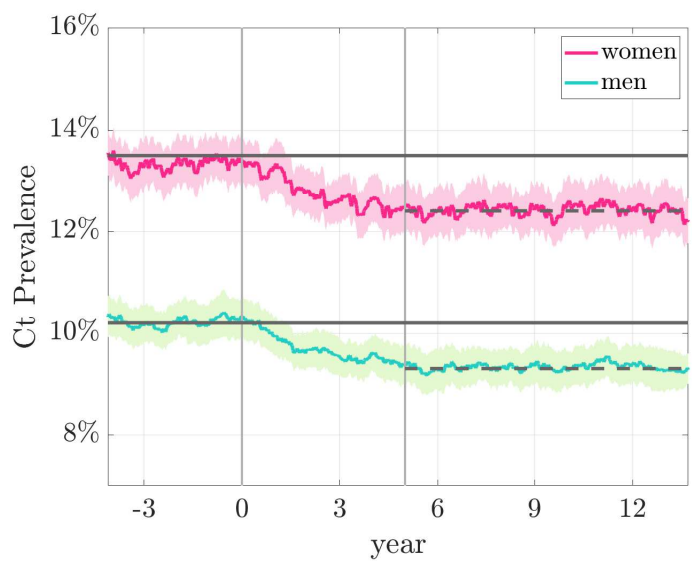}
\caption{Impact of male-screening program \CHK at the existing intervention level. The \CHK program starts at year zero, where the baseline Ct prevalence are $13\pt5$\% and $10\pt2$\% in women and men, respectively. Around year five, the Ct prevalence reach quasi-steady states, $12\pt4$\% and $9\pt3$\% in women and men, respectively. Thus, prevalence are reduced by $8\pt1$\% and $8\pt8$\% in women and men, respectively. The curves are the averages of 50 stochastic simulations. \label{fig:baseline}}
\end{figure}

We quantified the impact of \CHK mitigation at the current level of intervention intensity (shown in \cref{tab:parameters}). \Cref{fig:baseline} shows the change in Ct prevalence in both men and women after the launch of the \CHK screening for men at year zero with a balanced initial condition (\cref{sec:A_IC}). With the male-screening \CHK intervention, the Ct epidemic is controlled at a much lower prevalence.
In fact, near the quasi-steady state (around year five in \cref{fig:baseline}), our model predicts the following annual statistics from the program. Here shown with modelling scale followed by the target population in New Orleans in parentheses. The modelling scale is 5000 people, including 2304 men and 2696 women, and the target population (age range 15-24, sexually active) in New Orleans is 29,600 people, including 16,181 men and 13,419 women:

\begin{itemize}
\item[-] Each year, the program conducts in total 175 (1229) screenings, including 42 (295) from peer-recruited participants, and achieves 12 (84) treatments for index men (EIT) and 8 (56) treatment for partners of men (EPT).
\item[-] Among all the screened men found to be Ct+, the average number of partners within the past two months is $2\pt32$ (median = 2).
\item[-] Compared to the scenario without \CHKn, the program prevents 13 (91) cases in men and 41 (204) cases in women per year.
\item[-] Roughly, for each man screened, it could prevent $0\pt08$ case in men and $0\pt23$ case in women.
\end{itemize}

\subsection{Significance of the components of the intervention}
The sensitivity analysis identified the significant components in the intervention program. In the results presented below, we have considered the QOI to be $q =$ Ct prevalence and omit the upper index for the simplicity of the presentation. 

\begin{figure}[htbp]
\centering
\includegraphics[width=0.325\textwidth]{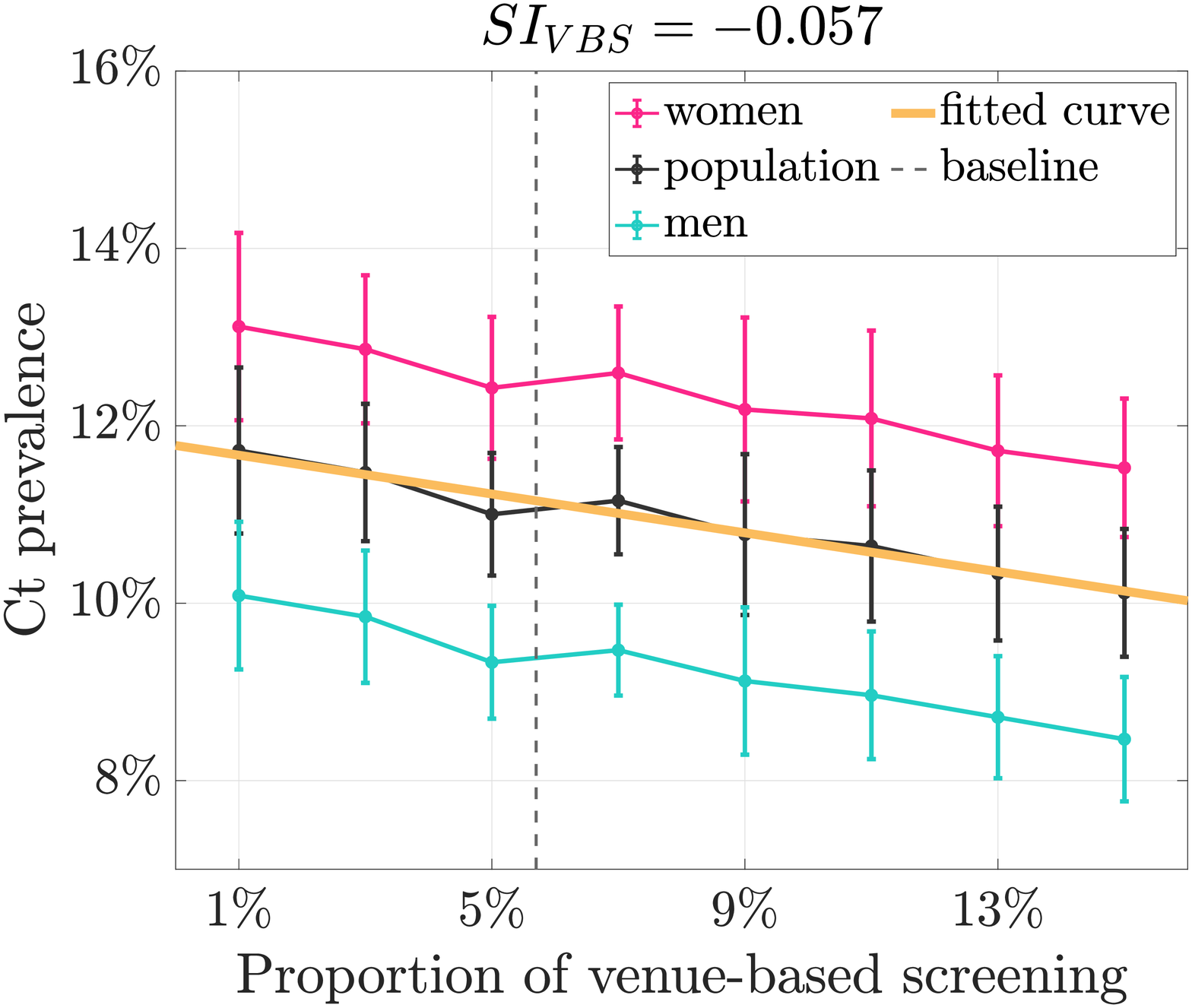}\hfill\includegraphics[width=0.325\textwidth]{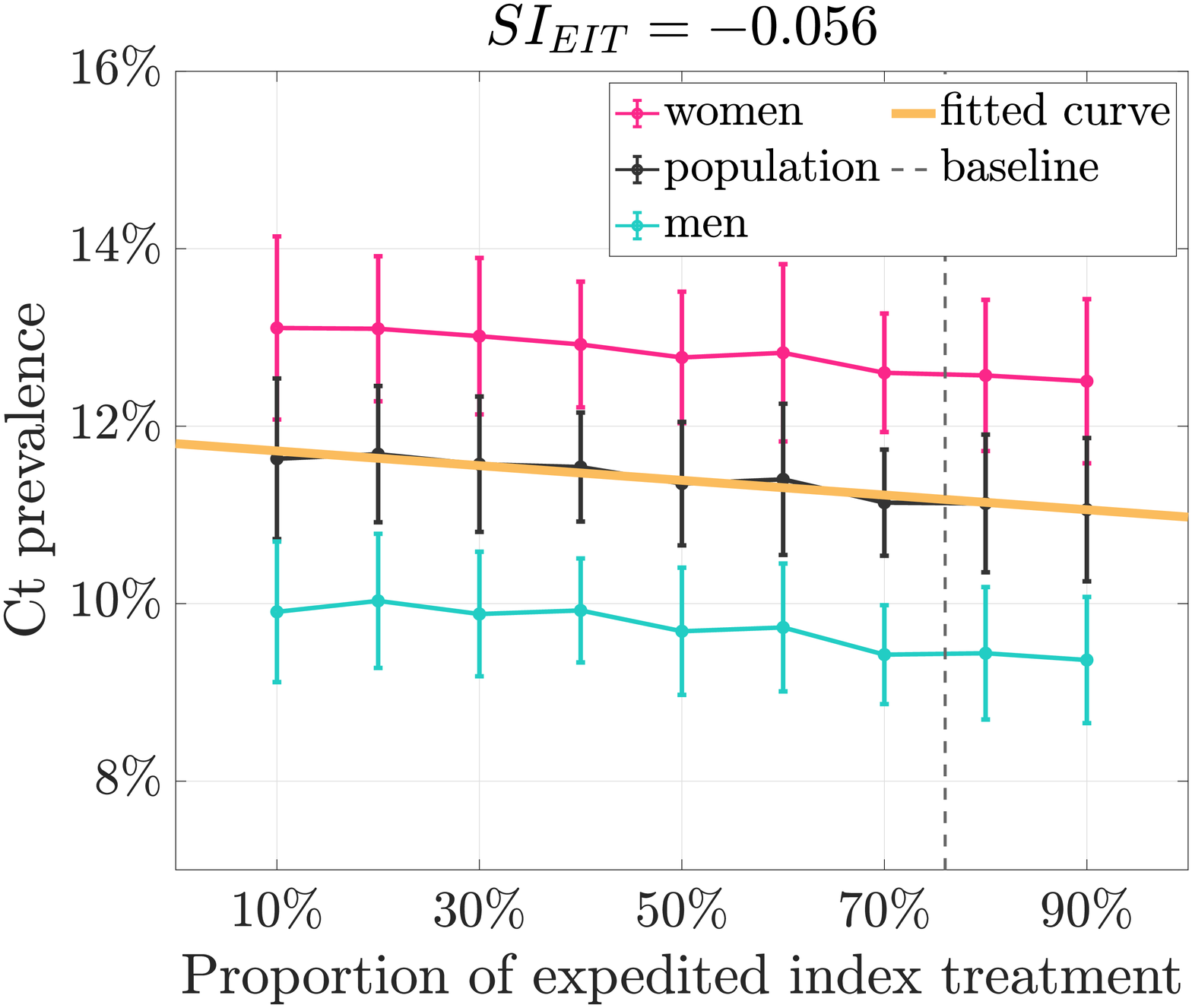}\hfill\includegraphics[width=0.325\textwidth]{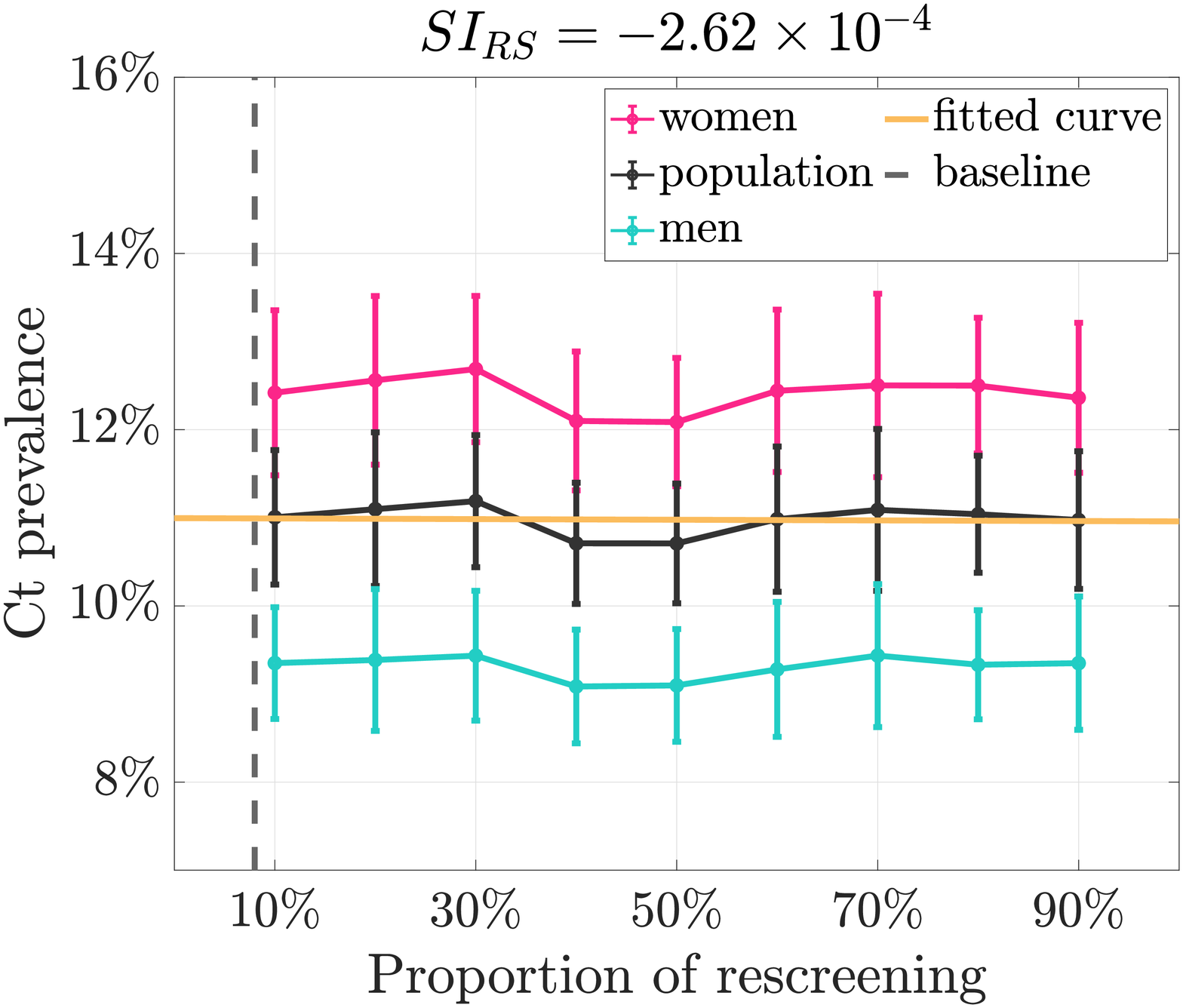}\\
\caption{Local sensitivity analysis on \CHK components (on the x-axis) against the Ct prevalence (y-axis) at the current level of intervention intensity. Titles are the corresponding local sensitivity indices $\mathcal{S}_p^q$ ($q=$ Ct prevalence, omitted for simplicity). The Ct prevalence are averaged over the time-frame year $4\pt5 \sim 5\pt5$ of 30 simulations. \label{fig:SA_1D}}
\end{figure}

\begin{figure}[htbp]
\centering
\includegraphics[width=0.35\textwidth]{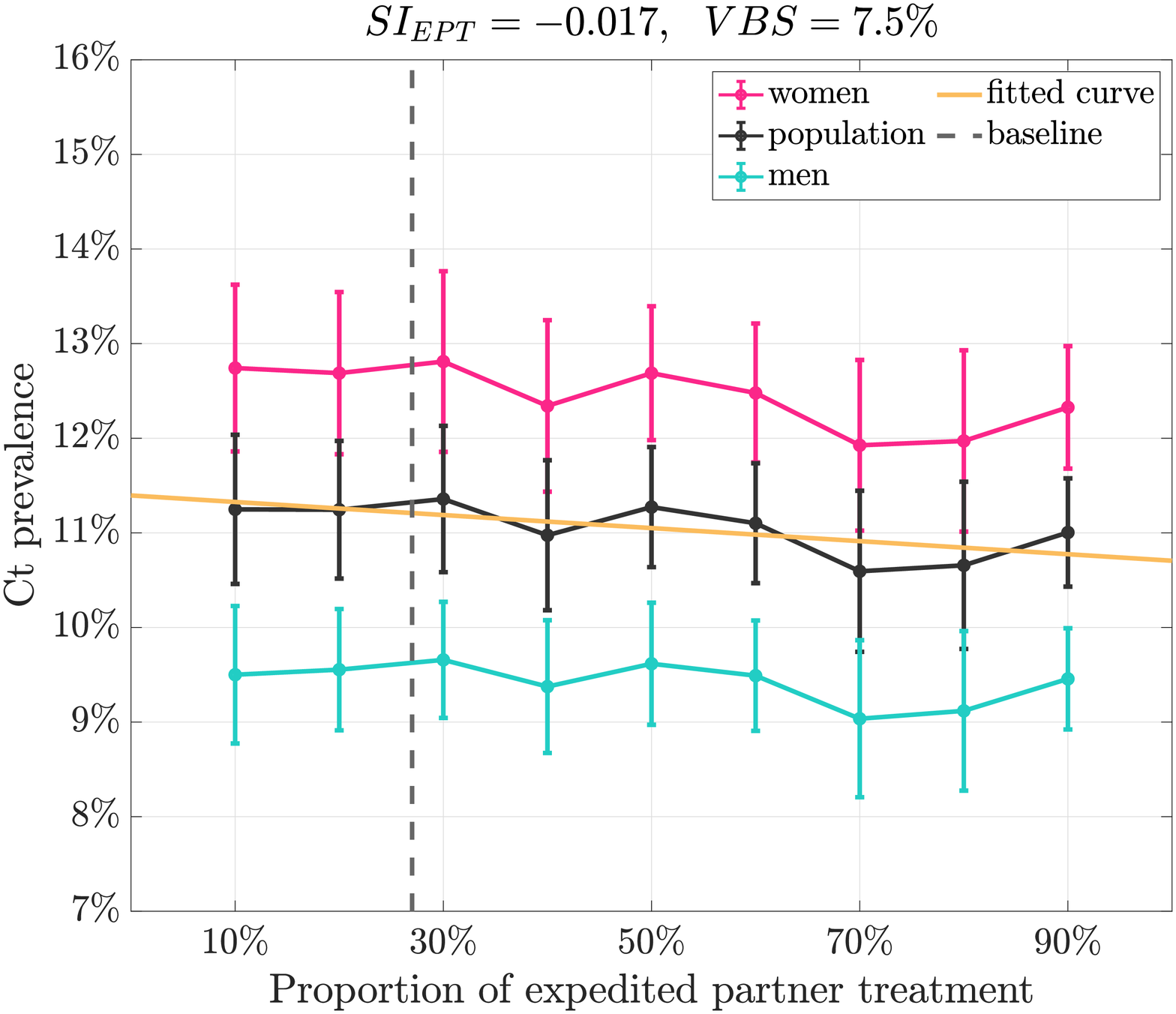}\hspace{1.5cm}\includegraphics[width=0.35\textwidth]{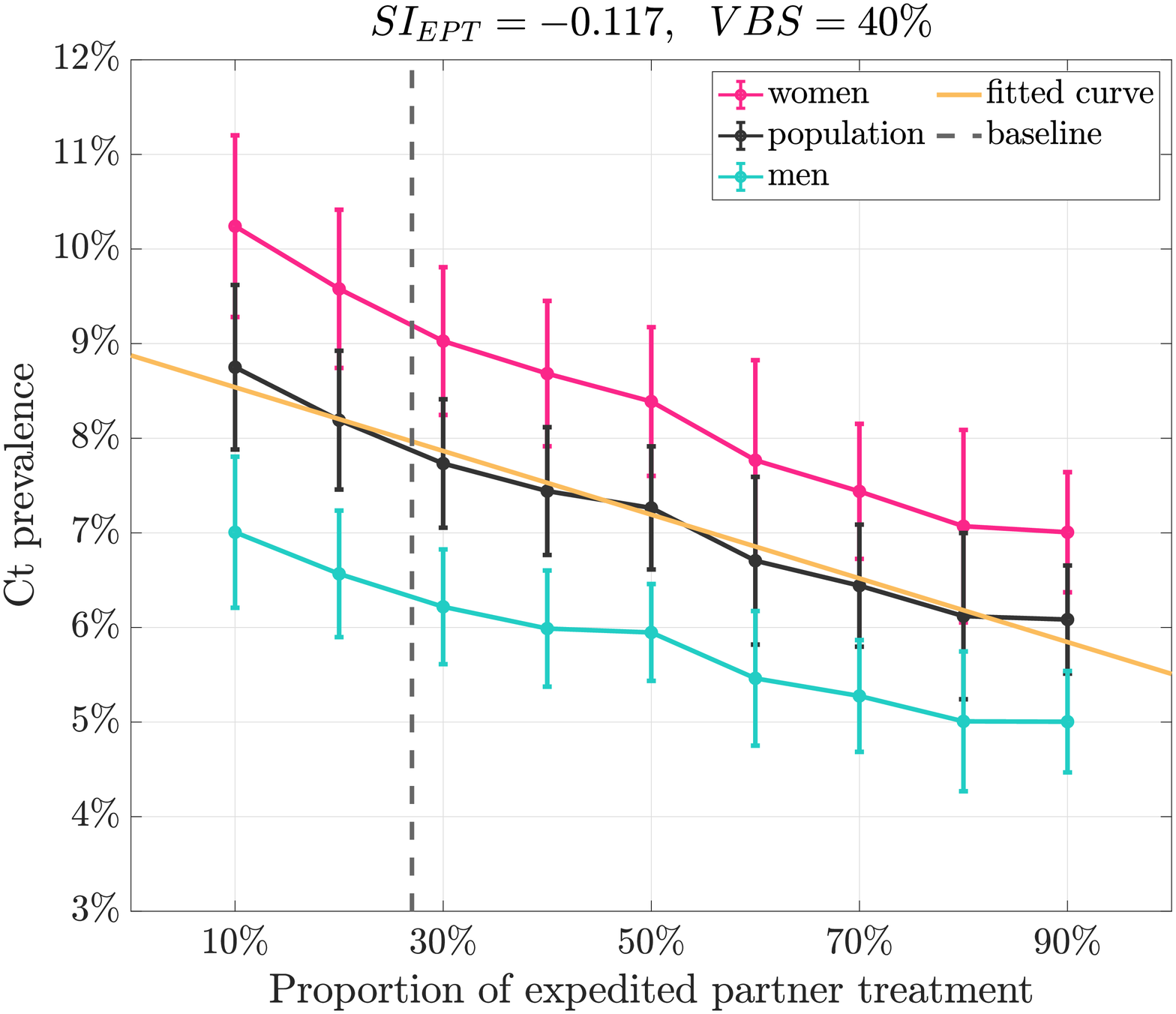}
\caption{Local sensitivity analysis on expedited partner treatment (on the x-axis) against the Ct prevalence (y-axis). \textbf{Left}: the analysis at the current level of \CHK intervention intensity, screening $7\pt5\%$ of the target male population. Together with the results in \cref{fig:SA_1D}, the significance of intervention components is ranked as venue-based screening (VBS) $\approx$ expedited index treatment (EIT) $>$ expedited partner treatment (EPT) $>$ rescreening. \textbf{Right}: the local sensitivity analysis at a much higher $40\%$ male-screening rate while fixing other intervention parameters. The magnitude of the sensitivity index is almost $10$ times larger ($-0\pt117$ vs. $-0\pt017$), which suggests that the partner treatment is more important in reducing prevalence when increasing the screening coverage in men. \label{fig:SA_1D2}}
\end{figure}

From \cref{fig:SA_1D} and \cref{fig:SA_1D2} (left), the Ct prevalence have an almost linear response to the intervention parameters, and the sensitivity at the current level of \CHK intensity is ranked as VBS $\approx$ EIT $>$ EPT $>$ rescreening. Moreover, when increasing the coverage of VBS from $7\pt5\%$ to $40\%$ (\cref{fig:SA_1D2}), the magnitude of the corresponding SI for EPT becomes $10$ times larger ($-0\pt117$ vs. $-0\pt017$), which suggests that EPT would be more effective in reducing the prevalence when having a high male-screening coverage.

\begin{figure}[htbp]
\centering
\includegraphics[width=0.6\textwidth]{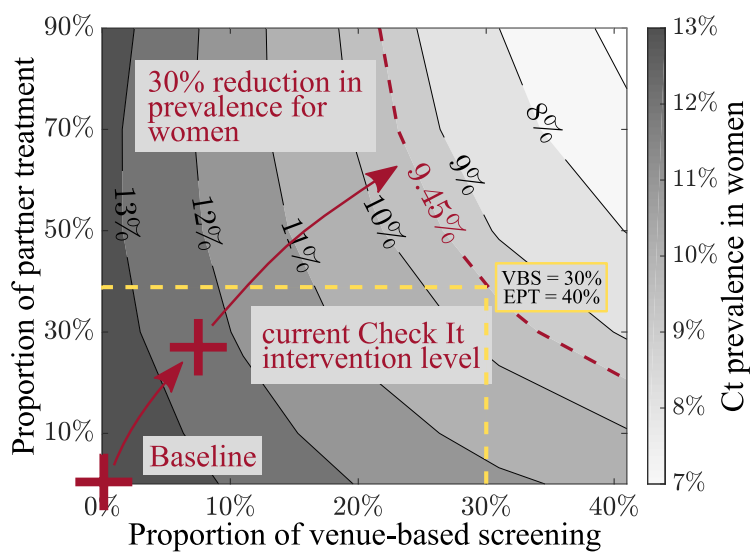}
\caption{Global sensitivity analysis of Ct prevalence in women (marked in contour lines) against two intervention parameters: venue-based screening (x-axis) and expedited partner treatment (y-axis) on a uniform $5\times5$ grid. At each grid point, the Ct prevalence is averaged over the time-frame year $4\pt5 \sim 5\pt5$ of 10 simulations, and the contour surface is smoothed by a least-square fit of a two-dimensional quadratic polynomial to the grid values. The baseline (VBS $= 0\%$, EPT $= 0\%$) and current intervention level (VBS $= 7\pt5\%$, EPT $= 27\%$) are marked in red crosses, which gives a $8\pt1\%$ reduction in women's prevalence. To achieve a $30\%$ reduction in women's Ct prevalence, the combined intervention levels needed are marked by the dashed contour line.  \label{fig:SA_2D}}
\end{figure}

We then conducted the global sensitivity analysis using the two most significant parameters from the local sensitivity analysis: VBS and EPT. The response plot for women's Ct prevalence (\cref{fig:SA_2D}) shows that the male-screening strategy has the potential to reduce the Ct prevalence in women substantially, and it predicts the effectiveness under different combinations of intervention intensities. For example, the model estimates the combination of VBS $= 30\%$ of the target men and EPT$= 40\%$ of their partners will give a $30\%$ reduction in Ct prevalence among women.


\section{Discussion}
Our model provided a framework for a public health team to ask ``what if'' questions that are hard to evaluate in the field. Findings of our analysis suggest that, in addition to the current female-screening intervention, the male-screening intervention, in conjunction with the EPT and rescreening, has the potential to mitigate the Ct prevalence among women. We estimated that by annually screening $7\pt5\%$ of the target male population (AA, sexually active young men, ageing 15 - 24), under the current implementation of \CHK program, the Ct prevalence is reduced by $8\pt1\%$ and $8\pt8\%$ in men and women, respectively. Our findings also show that as the coverage of male-screening increases, the EPT becomes more effective in reducing the Ct prevalence in women. 

We aimed to consider a comprehensive picture of the current Ct epidemic and control scenario, but we also recognised that mathematical models are a simplistic representation of the real world. The uncertainty and bias in the model parameters and model assumptions can affect the reliability of the quantitative predictions. Still, in our investigations, we found that the relative ranking of the responses in the sensitivity analysis and the trend in prediction pattern is robust and insensitive. 

Many of our model limitations are associated with the scope of the survey data sets collected from the target population. We described the sexual behaviour in this particular population through the sexual networks that are generated based on the survey data of \CHK and YGG in New Orleans. As detailed in \cref{sec:A11}, we matched the degree distribution and degree-degree distribution in the obtained sexual networks, which corresponds to a highly sexually active and high-risk population. 
We also considered the assortative mixing of partnerships on demographic traits, such as age, ethnicity, social groups, economic status, and geographic location, by embedding all the sexual networks in a grand social network, which describes the daily activities of residents in New Orleans. 

Although these data sets are from New Orleans, in many ways, they represent similar cities that have high Ct rates (urban, southern, and largely impoverished AA community). The same model could be applied for Ct transmission in other populations by adjusting sexual network and model parameters. The quantitative results obtained in this study can only be interpreted for other cities after carefully examining the differences between sexual behaviour and assortative mixing patterns in two populations.

Moreover, we implemented the dynamic sexual networks for simulating the epidemic over a long time. Every two months, we updated the sexual partnerships by replacing $50\%$ of the causal partner(s) by people from the same social contact network (details in \cref{sec:A13}). We chose a period of two months since it's the time frame covered by our \CHK survey data. Replacing $50\%$ of the casual partners every two months is a modelling assumption. Because of the lack of further data for the time between changing casual partners, we quantified how sensitive our results are to this assumption and compared the simulation results using different sets of parameters (\cref{sec:A13}). The numerical simulations show that the lower sexual mixing (longer dynamic period or lower fraction in partner update) can slow down the progress of the intervention, which leads to a longer time for Ct prevalence to settle into a new and lower quasi-steady state. However, the final (asymptotic) reduction in prevalence is relatively insensitive across different configurations that reflect different the level of sexual mixing in the population.

The model didn't consider the detailed immune responses for each individual during the infection process. Instead, the model was fit to the prevalence using the transmissibility parameters on the population level ($\beta^{m2w}$ and $\beta^{w2m}$). This gives the averaged and constant population-level approximations of the immunity levels for men and women. Also, we didn't incorporate the change in one's susceptibility to infection after the natural recovery, where partial protective immunity of some degree may develop from a genital infection \cite{batteiger2010Protective}. 

Another limitation is that our model assumed no sexual behaviour change, that is the dynamic partnership updating process preserved the total number of partnerships within each two-month time frame. We also assumed that the frequency of sexual contacts remains unchanged for the individuals before and after the \CHK intervention program. We will investigate the impact of behaviour change in future studies. These restrictions can be adjusted to predict the impact of behavior changes, such as reducing the number of partners, would have on Ct prevalence.  

We have modelled the epidemics and control on a closed population with no migration in/out of the network. That is, we did not consider the movement of people in/out the studying area, and no ageing effect: the same people stay in the age range 15-24, meaning they will not become too old and be removed from the eligible age range, and no young people will become 15 years old and come into the population. Thus, the model will only be a good approximation for a limited time, and it is not suitable for simulations over a long period. Our future research will improve the model by considering the ageing process in the population and migration, so we can better quantify the impact of a male-screening program in a more realistic setting.


\section*{Contributors}
ZQ contributed to the design of the mathematical model and data analysis, undertook numerical simulations and visualisation, and interpreted results. NS and MCC contributed to data collection and analysis and reviewed model design and interpretation. AA and CS reviewed the model design and interpretation. JH and PK contributed to the study, reviewed the model, data analysis and results interpretations, and oversaw and coordinated the investigation. ZQ wrote the first draft of the article. All authors contributed to the writing and review of the draft and approved final manuscripts.

\section*{Declaration of interests}
We declare no competing interests.

\section*{Acknowledgments}
This work was supported by the grants from the National Institutes of Health National Institute of Child Health and Human Development (R01HD086794) and the endowment for the Evelyn and John G. Phillips Distinguished Chair in Mathematics at Tulane University. The content is solely the responsibility of the authors and does not necessarily represent the official views of the National Institutes of Health.

\newpage
\appendix
\renewcommand{\theequation}{A.\arabic{equation}}
\setcounter{equation}{0}
\renewcommand{\thefigure}{A.\arabic{figure}}
\setcounter{figure}{0}
\renewcommand{\thetable}{A.\arabic{table}}
\setcounter{table}{0}
\renewcommand{\thesection}{A}
\setcounter{section}{0}
\section*{Appendix: Numerical implementations}

\subsection{Generation of dynamic sexual network \label{sec:A_network}}
We modelled the sexual behaviour of the target population through a series of simulated sexual networks, which were embedded in a grand social network. We characterised the sexual network by using the data inputs from two studies in New Orleans: ``You Geaux Girl!'' (YGG) and ``\CHKn'', where African-American women and men were enrolled to survey their sexual behaviour and complete an intervention process.  The social network was generated using Simfrastructure agent-based modelling and simulation system, which simulated a population of 150,000 in New Orleans.
\subsubsection{Degree and joint-degree distributions \label{sec:A11}}

We extracted the information of degree distributions (number of partners one reports) and joint-degree distributions (number of partners one thinks his/her partner has) for women and men from the YGG and \CHK data sets. These distributions were then used to generate sexual networks.   

\begin{figure}[ht!]
\centering
\includegraphics[width=0.45\textwidth]{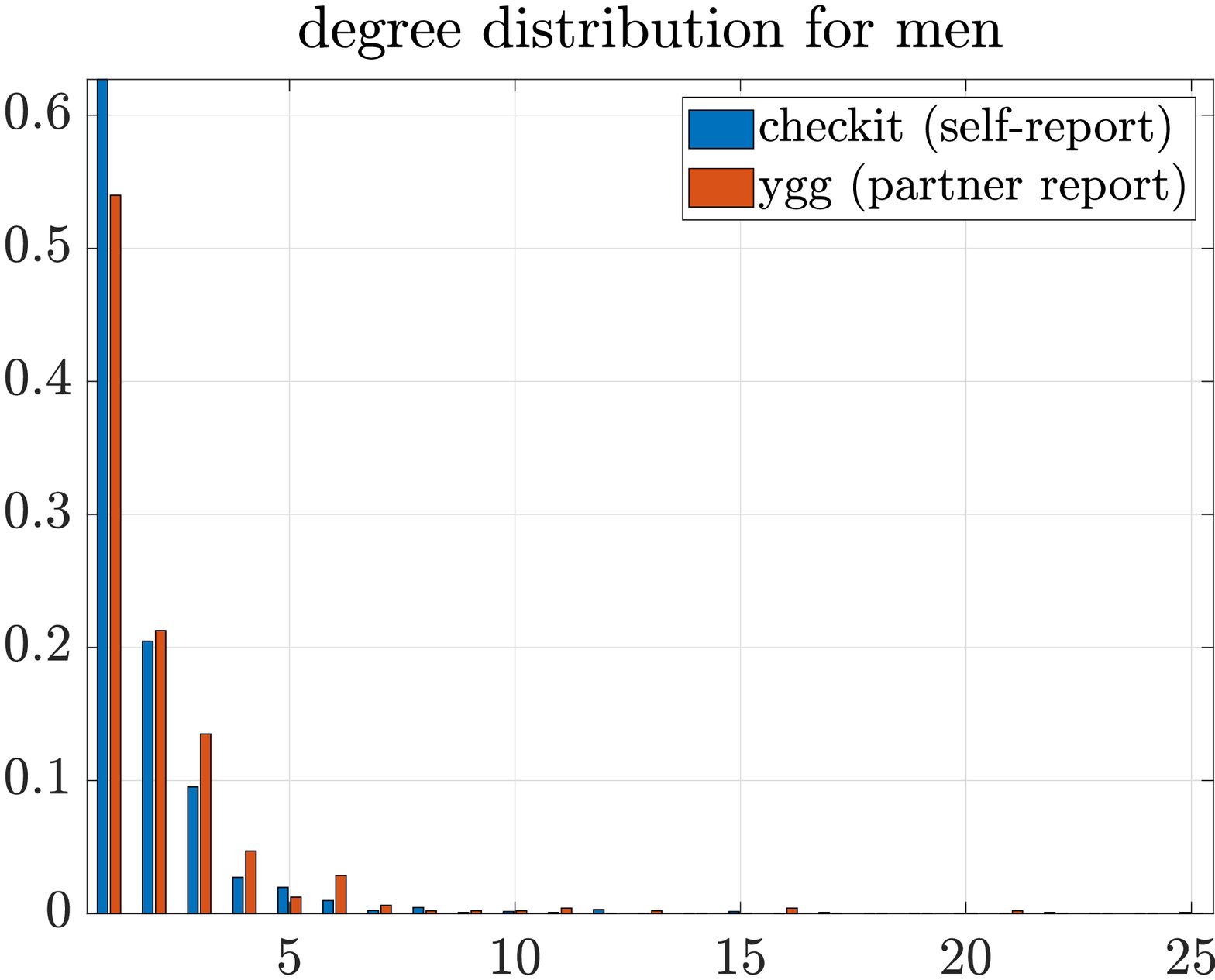}\hfill\includegraphics[width=0.45\textwidth]{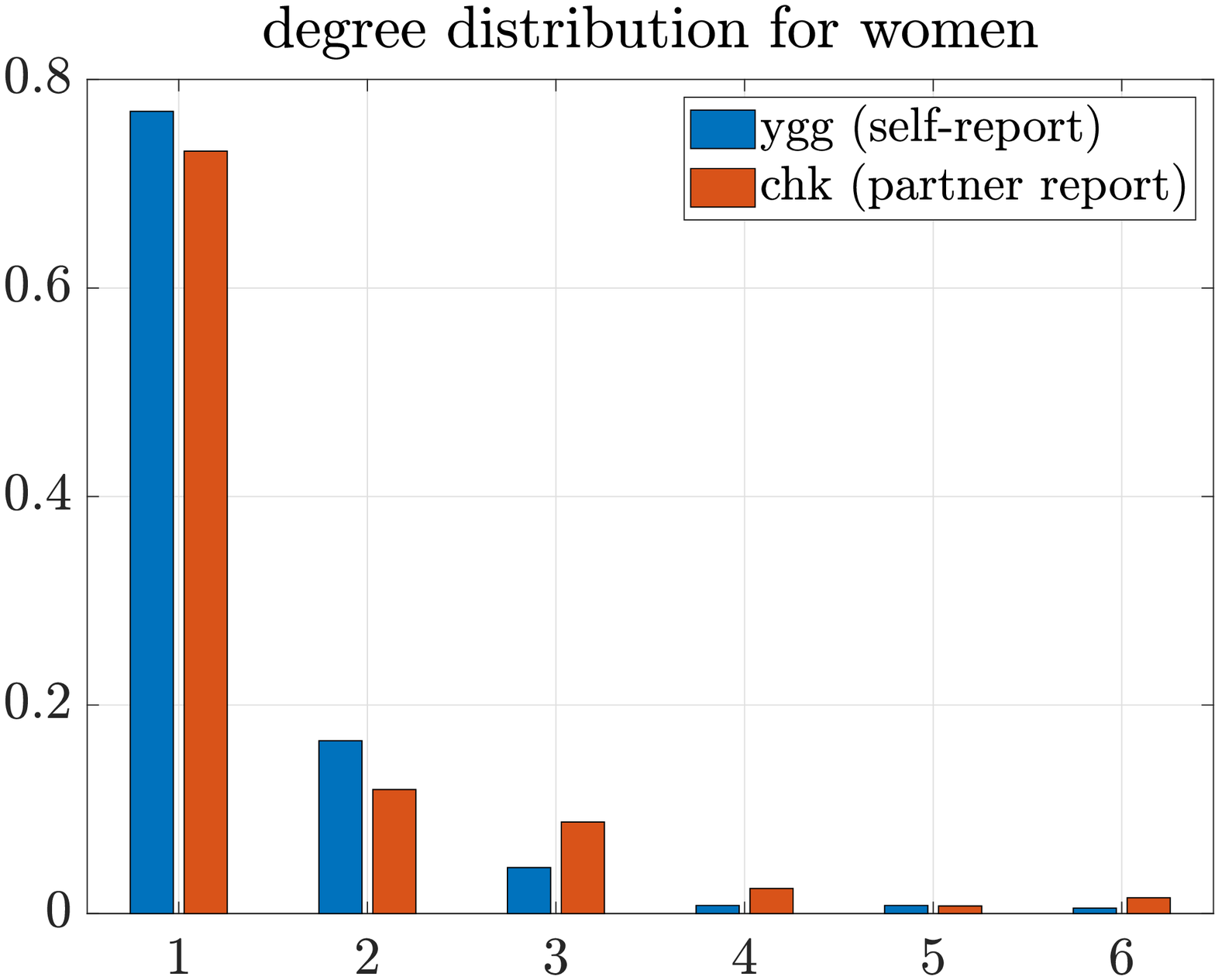}
\includegraphics[width=0.45\textwidth]{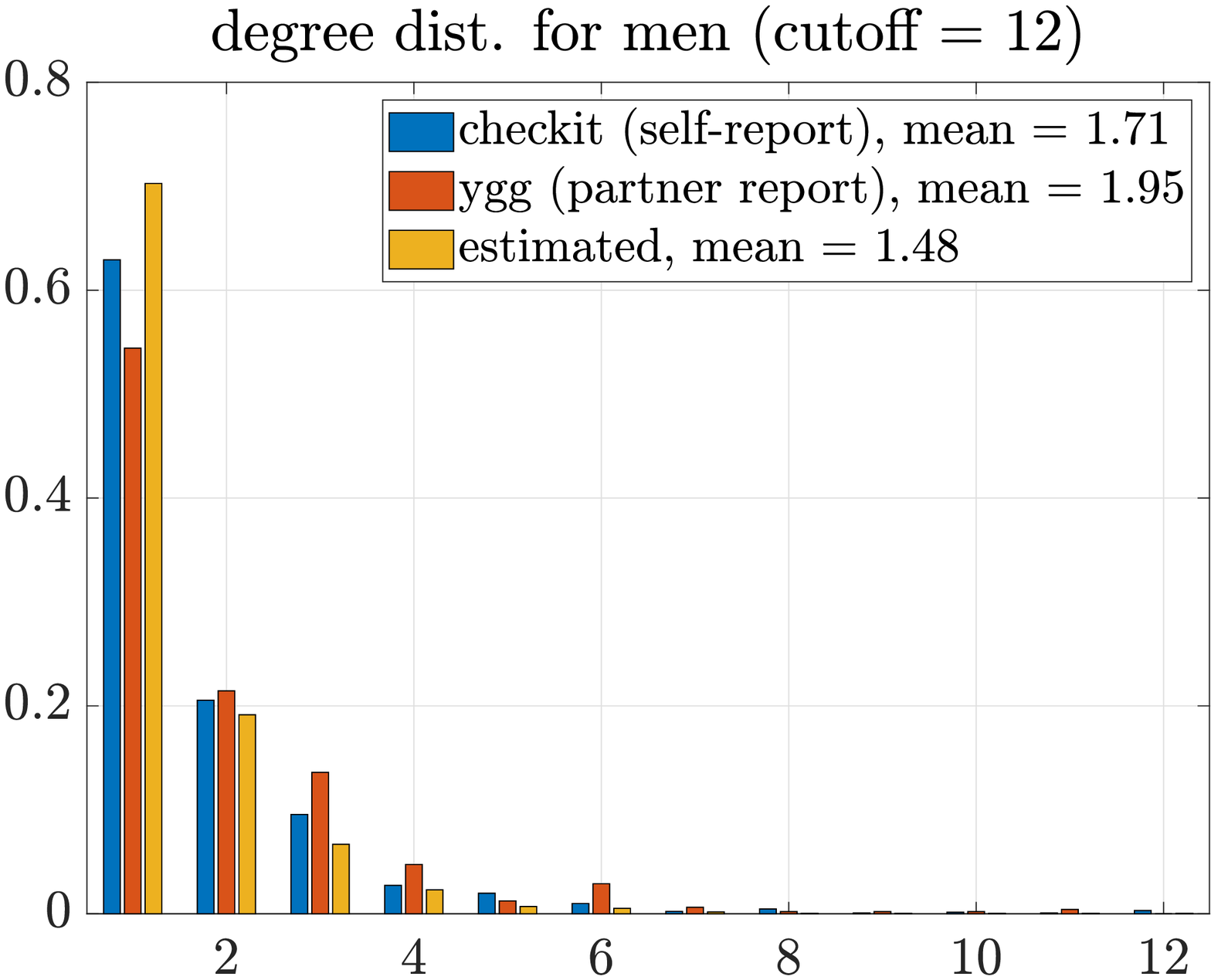}\hfill\includegraphics[width=0.45\textwidth]{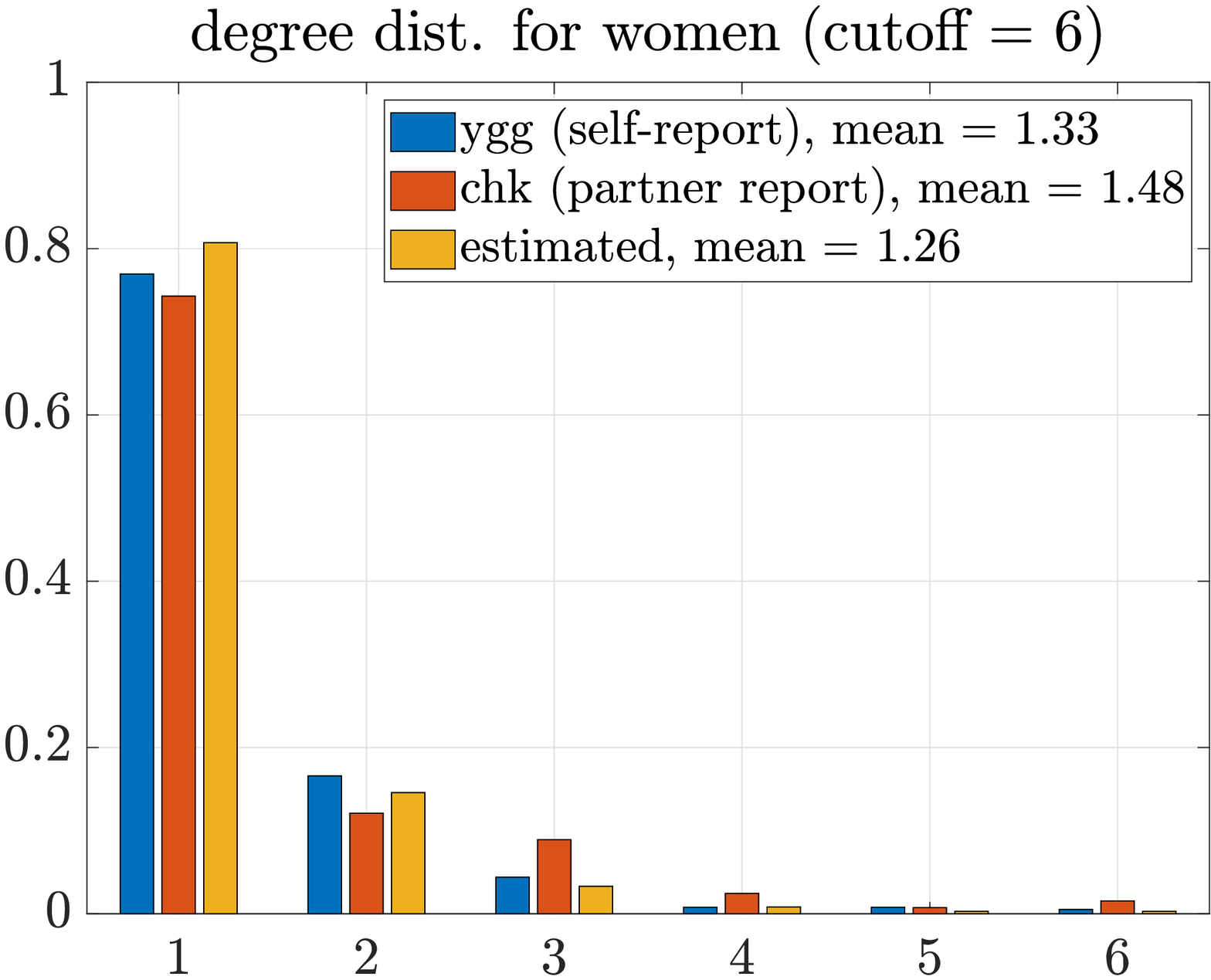}
\caption{Degree distributions for the number of partners for men and women. Blue bars: self-report data; Red bars: partner report data; Yellow bars: estimated distributions, which are used as model inputs to generate sexual networks. Top row: original degree distributions for men and women; Bottom row: processed degree distributions with cutoff at max degree = 12 for men and max degree = 6 for women, respectively.\label{fig:data_deg}} 
\end{figure}

Based on the degree distributions from the self-reported data (see \cref{fig:data_deg} first row), we truncated the distribution at a maximum degree = 12 for men and max degree = 6 for women. We then smoothed the extracted joint-degree distribution using LOESS with second degree polynomials, which was implemented by \verb|fit| (method = \verb|`loess'|, span = $0\pt25$) in MATLAB. Lastly, we normalised the distribution to unity. \Cref{tab:deg-deg} shows the estimate of the joint-degree table on a 5000 population size, where the $(i,j)$ entry in the table gives the total number of partnerships that exist between degree $i$ women and degree $j$ men on the sexual network. Figure in \cref{tab:deg-deg} shows the smoothed surface of the two-dimensional joint-degree distribution. 
We then recalculated the estimated degree distributions from the updated joint-degree distribution, and we compared it with the reported ones in the second row of \cref{fig:data_deg}. Overall, the estimated degree distributions still in good agreement with the ones from the self-report data.

\begin{table}[ht]
\hspace{-1cm}
\begin{minipage}[c]{0.6\textwidth}
\centering
\scalebox{0.8}{
\begin{tabular}[c]{@{}cc|cccccccccccc@{}}
\multicolumn{2}{c}{} & \multicolumn{12}{c}{\# of partners for men (degree of men)} \\
 &		&    \textbf{1} & \textbf{2} & \textbf{3} & \textbf{ 4} & \textbf{ 5} &\textbf{ 6}  &\textbf{ 7} &\textbf{ 8 }&\textbf{ 9} &\textbf{ 10 }& \textbf{ 11} & \textbf{ 12} \\
\midrule
\multirow{6}{1cm}{\centering{women degree}} &   \textbf{1} & 1102 & 608 & 273 & 109 & 42 & 29 & 15 & 6 & 5 & 4 & 6 & 6 \\
		&    \textbf{2} & 425  & 163 & 100 & 34  & 14 & 14 & 2  & 1 & 1 & 3 & 3 & 3 \\
		&    \textbf{3} & 77   & 57  & 52  & 32  & 17 & 6  & 2  & 1 & 2 & 2 & 0 & 1 \\
		&    \textbf{4} & 17   & 16  & 20  & 13  & 2  & 2  & 0  & 0 & 1 & 0 & 0 & 1 \\
		&    \textbf{5} & 0    & 3   & 14  & 0   & 3  & 7  & 2  & 0 & 0 & 0 & 0 & 1 \\
		&    \textbf{6} & 6    & 13  & 15  & 8   & 2  & 2  & 0  & 0 & 0 & 0 & 2 & 0
\end{tabular}}
\par\vspace{0pt}
\end{minipage}
\hfill
\begin{minipage}[c]{0.36\textwidth}
\centering
\includegraphics[width=1.1\linewidth]{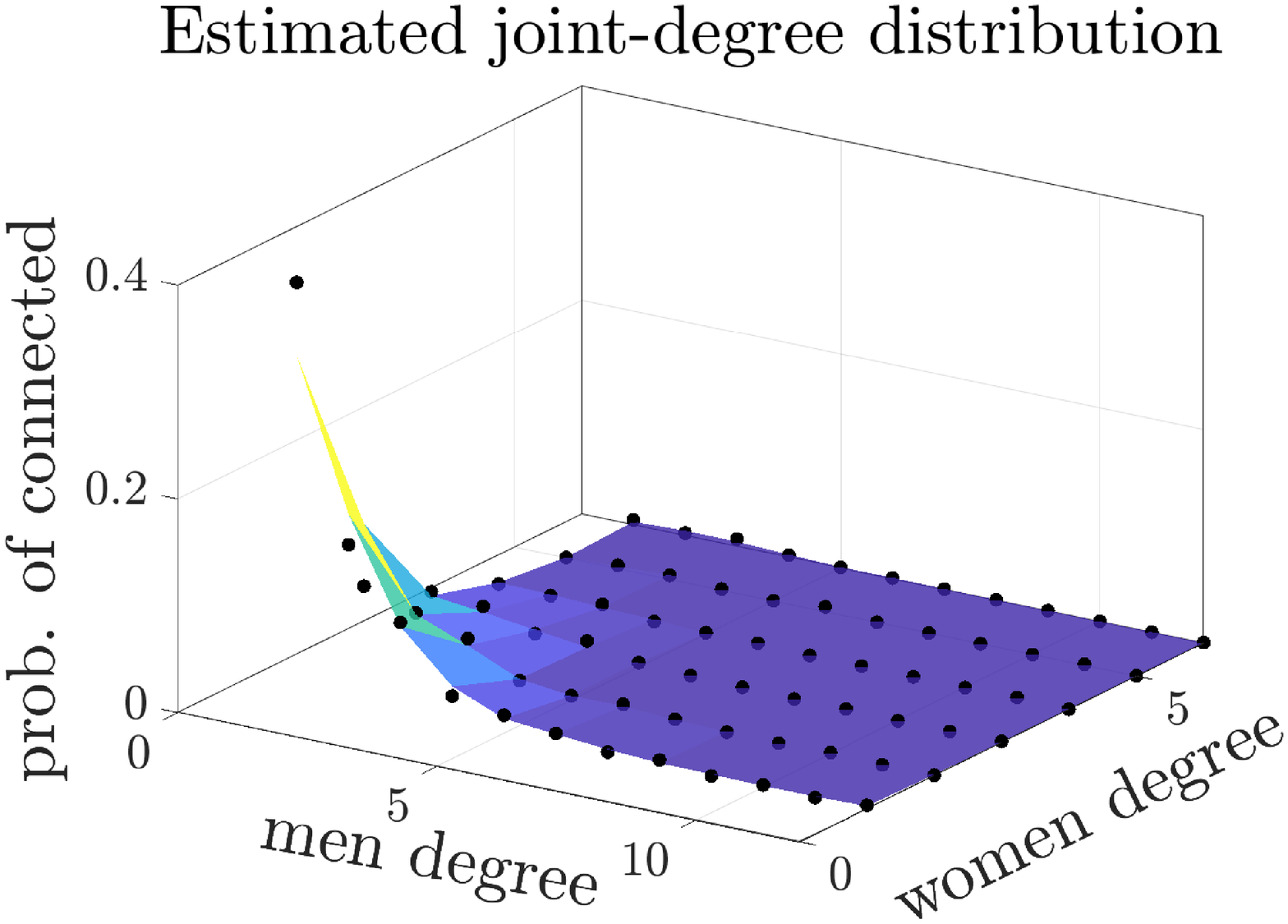}
\par\vspace{0pt}
\end{minipage}
\caption{\textbf{Left Table}: Joint-degree table for a population of size 5000. The entry at $(i,j)$ location represents the total number of partnerships between degree $i$ women and degree $j$ men in the sexual network. \textbf{Right Figure}:  Plot of the corresponding joint-degree probability distribution. \label{tab:deg-deg}}
\end{table}

\subsubsection{Primary and casual partnership \label{sec:A12}}
We categorised the partnership into primary (long-term) and casual (short-term), where the casual partners may be replaced every two months, which is the time frame that the \CHK survey asks about recent sexual behaviour, and the primary partners stay in the sexual network. 

For women, the YGG survey reported partner type as ``main'' and ``casual'', thus, we considered the main partner is our primary partner class. For men, the \CHK survey has several questions that characterise the relationship from different perspectives, including the duration of partnership, where they met (school, neighbourhood, club, dating site, etc), the best description of the relationship (girlfriend, wife, close friend, one-night stand), feeling committed or not and level of closeness. For \CHK data, there are in total 2052 partners for 1318 men, and the partnership types were assigned using the following ordered criteria:
\begin{enumerate}
\item If the relationship is best described by an ex-girlfriend (n=234), someone who I might want to have a relationship with (n=70), one-night stand (n=136), someone I paid to have sex with (n=10), internet hook up (n=14), or other (such as stranger, coworker and random answers, n=55), refuse to answer, or don't know on relationship type (n=137) $\rightarrow$ casual partner.
\item If the relationship is best described by: girlfriend (n=596), wife (n=65) $\rightarrow$ primary partner.
\item For other relationship types, including a good friend of mine, a friend with benefits, someone I have sex with but not necessarily a friend:
\begin{enumerate}
\item if consider oneself committed (n=54) $\rightarrow$ primary partner
\item If reporting refuse to answer or don't know on the duration of the sexual relationship (n=64), or times of vaginal sex in the past 2 months (n=15) $\rightarrow$ casual partner.
\item If times of vaginal sex $\ge 5$ (n=133) $\rightarrow$ primary partner.
\item If times of vaginal sex $\le 1$ (n=236) $\rightarrow$ casual partner.
\item If times of vaginal sex $ 2-4$:
\begin{enumerate}
\item If the duration of sexual relationship greater than six months (n=54) $\rightarrow$ primary partner.
\item In the remaining undefined partners: if the level of closeness or strength of your relationship (on a scale 1-10, 1 being not close or strong at all and 10 being extremely close) is greater than six (n=88) $\rightarrow$ primary partner.
\item In the remaining undefined partners: if meeting with the partner before first had sex with her at a club or other event and didn't know her before (n=3), meet her online through a dating site or social media (n=1) and other (party, work, college, etc, n=3) $\rightarrow$ casual partner.
\end{enumerate}
\end{enumerate}
\item Among all the remaining undefined partnerships, if the man 
\begin{enumerate}
\item already has exactly one primary partner (25 men have exactly one primary partner following the rules above), then consider all other partner(s) (n=26) $\rightarrow$ casual partner;
\item already has exactly two primary partners (four men have exactly two primary partners following the rules above), then consider all other partner(s) (n=6) $\rightarrow$ casual partner;
\item (no more than two primary partners defined for all the men in \CHK data)
\item has exactly one partner, who is undefined (n=3), both the level of closeness is less than two $\rightarrow$ casual partner;
\end{enumerate}  
\item all other undefined partners (n=49) $\rightarrow$ casual partner.
\end{enumerate}

We used our best guess to categorise each partnership based on the process described above, and we summarised the resulting distributions in \cref{tab:primary_partner}, which is not sensitive to the perturbation of the classification criteria. For each table, the $(i,j)$ entry gives the fraction of degree $i$ person who has exactly $j$ primary partner(s), and the last column is the accumulative probability for having at least one primary partner.

\begin{table}[ht]
\centering
\begin{tabular}{@{}ll|lllll@{}}
 & 	    & \multicolumn{5}{c}{\# of primary partners}  \\
 & 		& \textbf{1} & \textbf{2} & \textbf{3} & \textbf{4}  & $\ge$ \textbf{1}\\ 
\midrule
\multirow{12}{1.5cm}{\centering{degree of men}}  &\textbf{1}  & $0\pt66$ & 0    & 0    & 0    & $0\pt66$ \\
 &\textbf{2}  & $0\pt44$ & $0\pt15$ & 0    & 0    & $0\pt59$ \\
 &\textbf{3}  & $0\pt37$ & $0\pt23$ & $0\pt09$ & 0    & $0\pt69$ \\
 &\textbf{4}  & $0\pt36$ & $0\pt14$ & $0\pt11$ & $0\pt08$ & $0\pt69$ \\
 &\textbf{5}  & $0\pt15$ & $40\pt19$ & $0\pt08$ & $0\pt08$ & $0\pt5$  \\
 &\textbf{6}  & $0\pt46$ & $0\pt23$ & $0\pt08$ & 0    & $0\pt77$ \\
 &\textbf{7}  & 0    & $0\pt33$ & 0    & 0    & $0\pt33$ \\
 &\textbf{8}  & $0\pt33$ & $0\pt17$ & 0    & 0    & $0\pt5$  \\
 &\textbf{9}  & 1    & 0    & 0    & 0    & 1    \\
 &\textbf{10} & 0    & 0    & 0    & 0    & 0    \\
 &\textbf{11} & 1    & 0    & 0    & 0    & 1    \\
 &\textbf{12} & $0\pt75$ & $0\pt25$ & 0    & 0    & 1   
\end{tabular}\hfill
\begin{tabular}{@{}ll|llll@{}}
 & 	    & \multicolumn{4}{c}{\# of primary partners}  \\
 & 		& \textbf{1} & \textbf{2} & \textbf{3} & $\ge$ \textbf{1} \\  
\midrule 
\multirow{6}{1.8cm}{\centering{degree of women}} 	
 & \textbf{1} & $0\pt9$ & 0    & 0    & $0\pt9$ \\
 &\textbf{2} & $0\pt7$ & $0\pt08$ & 0    & $0\pt77$ \\
 &\textbf{3} & $0\pt8$ & $0\pt1$ & 0 & $0\pt9$ \\
 &\textbf{4} & 0 & 1 & 0    & 1 \\
 &\textbf{5} & 1    & 0    & 0    & 1    \\
 &\textbf{6} & 1    & 0    & 0    & 1   
\end{tabular}
\caption{Distributions of the number of primary partners for men (left) and women (right): the $(i,j)$ entry gives the probability that a degree $i$ man/women has (self-report) $j$ primary partners. The last column gives the accumulative probabilities for having at least a primary partner. \label{tab:primary_partner}}
\end{table}
\subsubsection{Dynamic sexual networks}\label{sec:A13}

Given the distributions estimated in \cref{sec:A11} and \cref{sec:A12}, we could generate a sexual network embedded in the social network, that is on average, the individual has a fraction $p$ of the casual partners from his/her connected social network, and a fraction $1-p$ are chosen randomly from the rest of the 5000 population. Moreover, to simulate the partner changing process in the population, we implemented dynamic sexual networks: every two months, with $50\%$ probability, the casual partners were replaced by either individuals from his/her social network or the rest of the population. The fraction $p$ was estimated from the \CHK survey data, which is $p=0\pt82$. 

\begin{figure}[ht]
\centering
\includegraphics[width=0.33\textwidth]{run_baseline_chk}\hfill\includegraphics[width=0.33\textwidth]{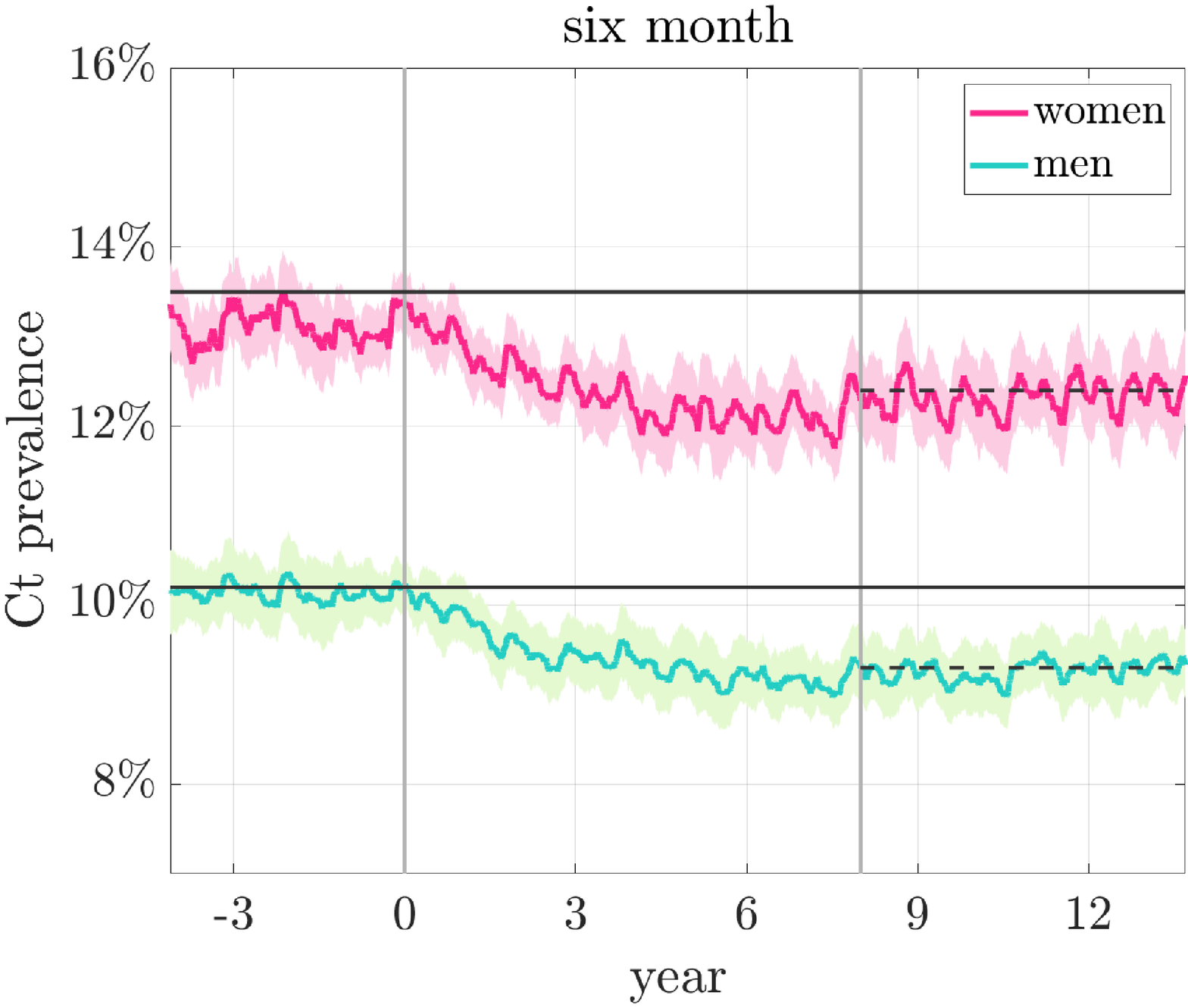}\hfill\includegraphics[width=0.33\textwidth]{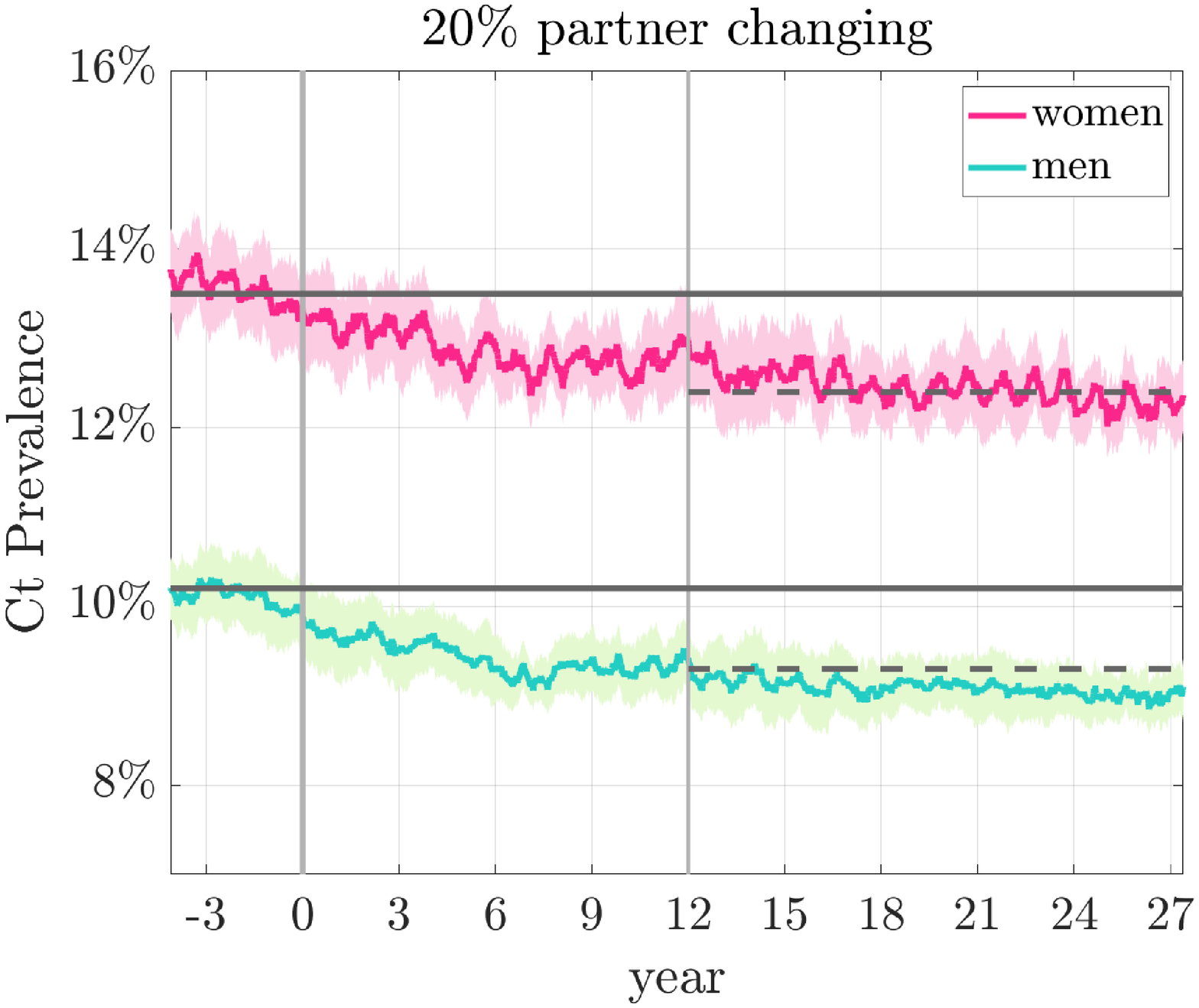} 
\caption{Dynamic sexual networks with different configurations. Left: configuration used in the model, ($50\%$, 2 months, $\beta^{m2w} = 0.3, \beta^{w2m} = 0.1$); Middle: ($50\%$, six month, $\beta^{m2w} = 0.34, \beta^{w2m} = 0.113$); right: ($20\%$, two month, $\beta^{m2w} = 0.34, \beta^{w2m} = 0.107$). It takes five, eight, and twelve years for Ct prevalence to achieve a lower quasi-steady state, respectively. \label{fig:dynamic}}
\end{figure}

We used two months for the partnership updates, since this is the time frame covered by the \CHK survey. The fraction $50\%$ is an assumption, not backed by survey data. To investigate the impact of these two parameters on the prediction, we ran simulations using two sets of different parameters, as shown in \cref{fig:dynamic}. For all the simulations, we have calibrated the model using transmission parameters $\beta^{m2w}$ and $\beta^{w2m}$ to the current Ct prevalence in men and women, as described in \cref{sec:A_IC}. By increasing the dynamic period and decreasing the fraction of causal partner replacement, the sexual mixing in the population is lowered. This also results in higher calibrated transmission parameters to match the same baseline Ct prevalence. Also, due to the lower sexual mixing level, it takes much longer for the program intervention to bring the epidemic down to a lower quasi-steady state. Overall, the effectiveness of the intervention, measured by the amount of (asymptotic) reduction in prevalence, is comparable to the baseline configuration and is relatively insensitive across different levels of sexual mixing in the population.

\subsection{Chlamydia transmission SIS model over a dynamic sexual network \label{sec:A_SIS}}
We updated the changes of infection statuses for the individuals daily, following the SIS framework, where each individual is either susceptible (S) or infected (I). 
\subsubsection{Force of infection: $S \rightarrow I$}
On each day, we modelled the force of infection, the ability for the infected individuals to spread Ct to their susceptible partners as follows:
\begin{equation*}
\resizebox{\textwidth}{!}{
$
\begin{aligned}
&\text{Trasmission}\\
&~~\text{or not} 
\end{aligned}
= 
\left[
\begin{aligned}
&~~\text{Prob. of having a sexual }\\
&\text{contact per day per partner}
\end{aligned}
\right]\cdot
\left[
\begin{aligned}
&\text{Prob. of transmission}\\
&~~~~~~~\text{per contact}
\end{aligned}
\times
\left(
1-
\begin{aligned}
&\text{Prob. of effective}\\
&~~\text{condom use}
\end{aligned}
\right)
\right],$}
\end{equation*}
where each bracket gives a zero or one value with the specified probability contained inside.

From the datasets, we summarised the probability of having a sexual contact per day per partner for men and women with different degrees in \cref{tab:contact_men_women}. This probability depends on both how many partners the individual has (degree of men/women) and the type of the partnership (primary or casual). In general, the probability of having contact with the primary partner is higher than with the casual one.

\begin{table}[ht!]
\centering
\begin{tabular}{@{}cc|ccc@{}}
 & 	    	 & \multicolumn{3}{c}{Type of partnership}  \\
 &			 & Primary     & Casual      & Mix-type       \\
 \midrule
 \multirow{12}{1.3cm}{\centering degree of men (\CHKn)}  
 & \textbf{1}  & $0\pt14$ & $0\pt06$ & $0\pt11$ \\
 & \textbf{2}  & $0\pt11$ & $0\pt06$ & $0\pt08$\\
 & \textbf{3}  & $0\pt12$ & $0\pt05$ & $0\pt08$ \\
 & \textbf{4}  & $0\pt18$ & $0\pt06$ & $0\pt11$ \\
 & \textbf{5}  & $0\pt15$ & $0\pt04$ & $0\pt07$ \\
 & \textbf{6}  & $0\pt07$ & $0\pt03$ & $0\pt05$ \\
 & \textbf{7}  & $0\pt08$ & $0\pt03$ & $0\pt04$ \\
 & \textbf{8}  & $0\pt04$ & $0\pt03$ & $0\pt03$ \\
 & \textbf{10} & $0\pt03$ & $0\pt02$ & $0\pt02$ \\
 & \textbf{11} & $0\pt03$ & $0\pt02$ & $0\pt02$ \\
 & \textbf{12} & $0\pt07$ & $0\pt06$ & $0\pt06$
\end{tabular}
\hfill
\begin{tabular}{@{}ll|lll@{}}
 & 	    	 & \multicolumn{3}{c}{Type of partnership} \\
 &  		 & Primary     & Casual      & Mix-type       \\
 \midrule
 \multirow{6}{1.7cm}{\centering degree of women (YGG)}  
 & \textbf{1} & $0\pt12$ & $0\pt02$ & $0\pt11$ \\
 & \textbf{2} & $0\pt07$ & $0\pt05$ & $0\pt06$ \\
 & \textbf{3} & $0\pt08$ & $0\pt02$ & $0\pt04$ \\
 & \textbf{4} & $0\pt02$ & $0\pt03$ & $0\pt03$ \\
 & \textbf{5} & $0\pt01$ & $0\pt01$ & $0\pt01$ \\
 & \textbf{6} & $0\pt02$ & $0\pt02$ & $0\pt02$
\end{tabular}
\caption{Probability of having contact per partner per day. For both tables, entries in row $i$ give the probability that a degree $i$ person has sexual contact with a primary (column one) or casual partner (column two), respectively. The last column gives the average probability of contact regardless of the partnership. \label{tab:contact_men_women}}
\end{table}

The probability of effective condom use depends on two factors, the probability of condom use and the probability of condom-use failure. From the data sets, we estimated that the values for condom use for primary and casual partners are $c_p=0\pt54$ and $c_c = 0\pt66$, respectively. The condom use failure rate was fixed as $c_\epsilon=0\pt1$.

In the case of asymmetric primary-casual relationship, we modelled both the condom use probability and the contact frequency as a compromise between the couple, and we took the harmonic average of the values from two sides. 

Lastly, the probability of transmission per contact ($\beta^{m2w}$ and $\beta^{w2m}$) can have a wide range in different scenarios. We considered the maximum likelihood estimates by solving a nonlinear least square optimisation problem and matching the Ct prevalence to the current prevalence in New Orleans ($10\pt2$\% in men and $13\pt5$\% in women).

\subsubsection{Temporary immunity: $I \rightarrow S$}
Meanwhile, the infected individual could recover (without lasting immunity) and become susceptible again. The recovery could be due to either the natural clearance of the pathogen or medical treatment. We modelled the time to recovery as exponential distributions with different means (see the baseline values in \cref{tab:parameters}). We also assumed that no one recovers naturally within the first three months of infection, which was incorporated as a shift in the corresponding distribution.

Moreover, both natural and treated recovering process could be interrupted and start over again if the individual has an infectious contact with the partner before fully recovered.  The natural recovery process could be updated as a treated recovery process once the individual gets treatment, either from the enrolment in the \CHK program (men), annual screening (women), or screening due to symptomatic infection.

\subsection{Model initialisation - balanced initial condition \label{sec:A_IC}}
\begin{figure}[htbp]
\centering
\includegraphics[width=0.7\textwidth]{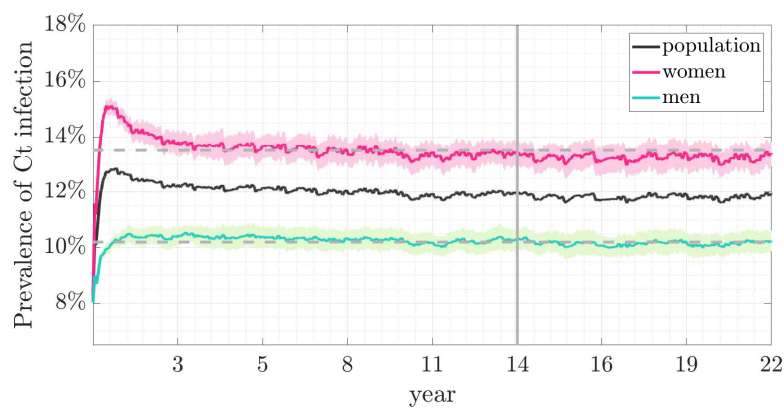}
\caption{Generating balanced initial condition: start the initial infection in 8\% of men and 8\% of women with the highest sexual degree, and a quasi-steady-state is achieved around year 14. We then reset the time to be zero and use this infection status as the initial condition for the rest of the simulations. The background lighter curves are the one standard deviation bands for 50 stochastic simulations, and the thicker curves in the middle are the mean of the simulations. The standard deviation of the prevalence is $0\pt0092$ and $0\pt0075$ for women and men, respectively. We calibrate the model to fit the current prevalence from data, which is $10\pt2$\% in men and $13\pt5$\% in women.  \label{fig:balanced_IC}} 
\end{figure}

We initiated the model to represent the current baseline Ct epidemic in New Orleans. Since the distribution of the current infected population depends on the history of the Ct epidemic, the initial infections are distributed across the sexual network as they would be as part of an emerging epidemic (\cref{fig:balanced_IC}). To identify a physically relevant initial condition, we started by an epidemic infecting a small fraction of the population at $t=0$, and let the infection be internally redistributed in time until a quasi-steady-state is achieved. We started with $8\%$ of infections, and for most situations, the simulations took about fourteen years. We then reset the time to be zero and use this infection status as the initial condition for the rest of the simulations. We have infected 8\% of the men and women with most sexual partners initially to speed up the process, as they are more likely to be infected in a balanced scenario.  	

\subsection{Modelling annual screening for women\label{sec:A_Female}}
The Centers for Disease Control and Prevention recommends annual Ct screening for all sexually active women younger than 25 as well as older women with risk factors, which is a standard preventive strategy to identify most of the asymptomatic infections in women. We considered this intervention strategy as part of the baseline scenario and implemented a simple model to simulate the women's annual screening process in practice. 

We assumed that a fraction $\sigma_a^w$ of the women in the target population receives annual Ct screening. In reality, most women return for screening following an annual routine schedule (regular annual screening), and there is also a small fraction of women receive screening on a casual basis (opportunistic screening). For the former case, we pre-generated an annual screening schedule for the regular screening group. For the latter case, we randomly sampled the opportunistic screening group from the rest of the women.

\begin{figure}[htbp]
\centering
\includegraphics[width=0.48\textwidth]{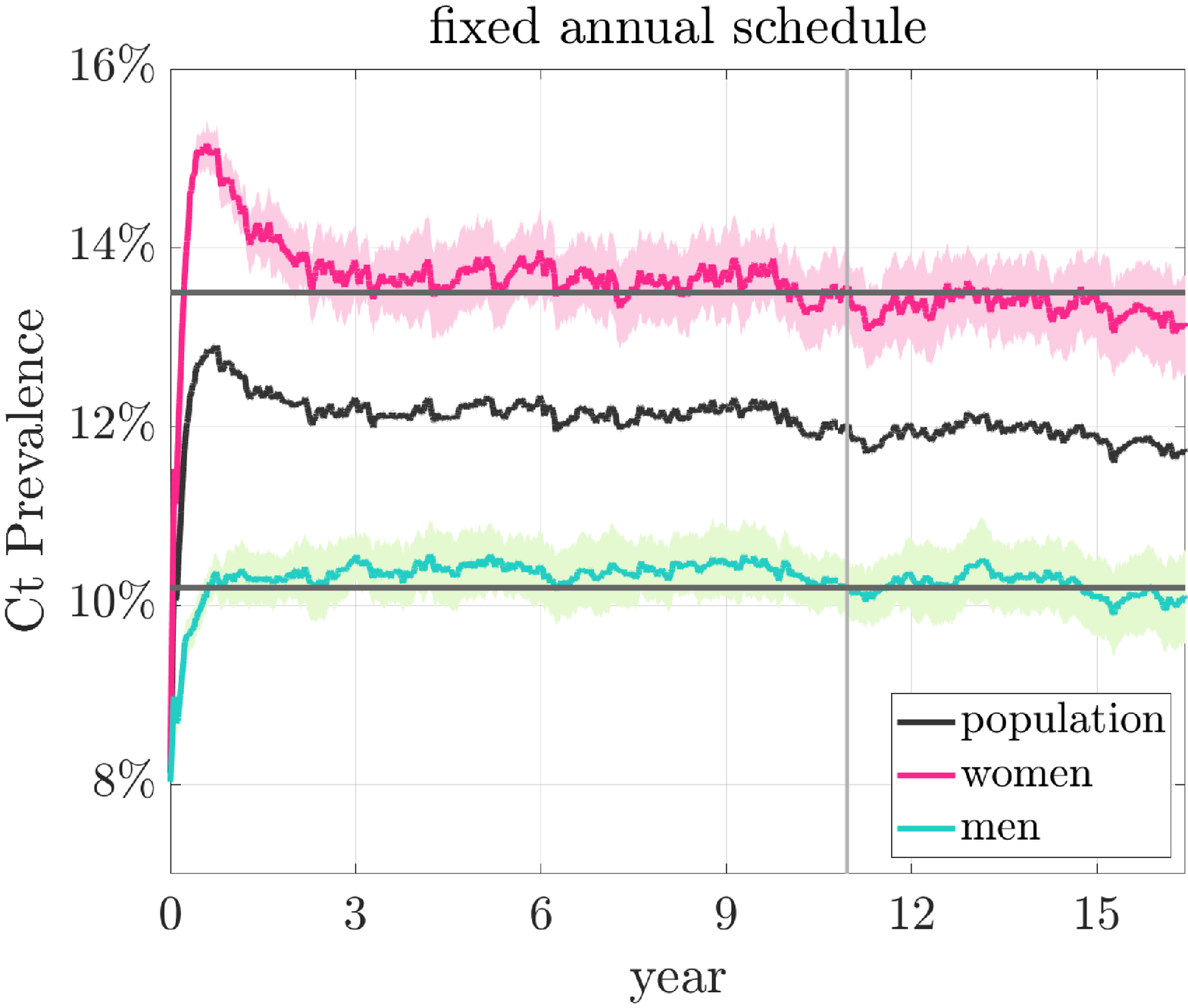}\includegraphics[width=0.48\textwidth]{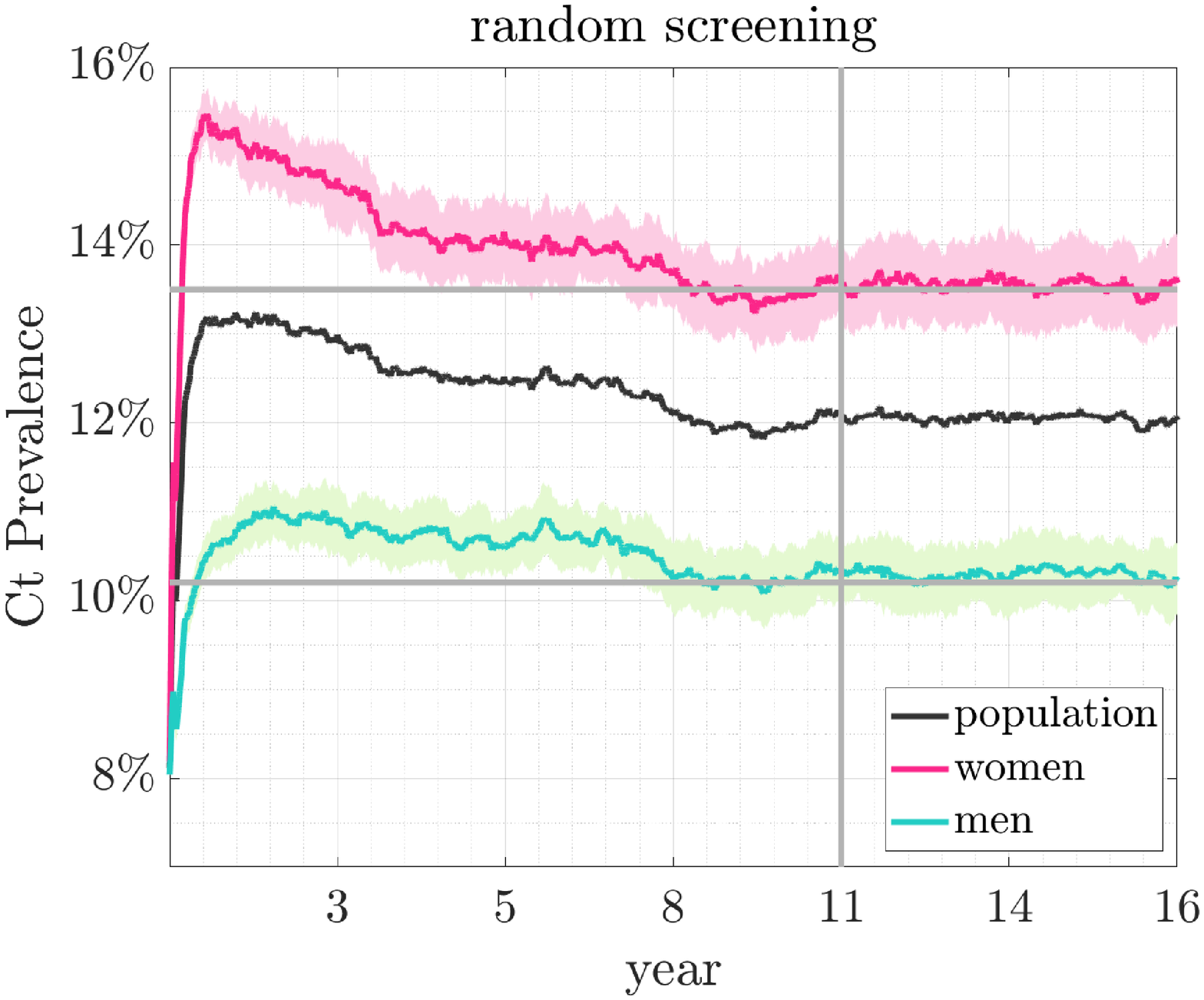}
\caption{Modelling the annual Ct screening as a preventive strategy in women under two scenarios: a constant population gets regular annual screenings following a fixed schedule (left), or randomly screen the same number of women but different populations from year to year  (right). The Ct prevalence for the random screening case converges much slower to a quasi-steady state than the fixed annual schedule case.\label{fig:IC3}}
\end{figure}

\Cref{fig:IC3} presents the simulations at the initial calibration stage for two extreme scenarios: all the screened population belongs to one group: either the regular screening group (left plot) or the opportunistic screening group (right plot). Two cases show similar trends and prevalence at the quasi-steady state. We observed that the random screening case has a significantly slower convergence to the quasi-steady state. The slower convergence is a result of random screening targets different populations from year to year rather than a fixed population and, therefore, it takes a longer time to converge to a stable status. In our model, since there is no further information on the proportion of the population within each group, we assumed that all the screenings belong to the regular screening group. Moreover, since we only studied the Ct scenario after the quasi-steady state is achieved, the difference at the initial convergence speed does not affect our conclusion.

\newpage
\bibliographystyle{vancouver2}
\bibliography{Chlamydia_checkit}

\begin{thebibliography}{30}

\bibitem[{{Centers for Disease Control and Prevention}(2019)}]{CDC2019Sexually}
{Centers for Disease Control and Prevention}.
\newblock Sexually {{Transmitted Disease Surveillance}} 2018.
\newblock Atlanta: US Department of Health and Human Services.
  2019;DOI:10.15620/cdc.79370.

\bibitem[{Hillis and Wasserheit(1996)Hillis, Susan D. and Wasserheit, Judith
  N.}]{hillis1996Screening}
Hillis SD, Wasserheit JN.
\newblock Screening for chlamydia -- a key to the prevention of pelvic
  inflammatory disease.
\newblock N Engl J Med. 1996;334(21):1399--1401.

\bibitem[{Ward and R{\"o}nn(2010)Ward, Helen and R{\"o}nn,
  Minttu}]{ward2010contribution}
Ward H, R{\"o}nn M.
\newblock Contribution of sexually transmitted infections to the sexual
  transmission of {{HIV}}.
\newblock Curr Opin HIV AIDS. 2010 Jul;5(4):305--310.

\bibitem[{{Centers for Disease Control and Prevention}(2007)}]{report2007male}
{Centers for Disease Control and Prevention}.
\newblock Male Chlamydia screening consultation, Atlanta, Georgia, March 28-29,
  2006, Meeting report.
\newblock National Center for HIV/AIDS, Viral Hepatitis, STD, and TB
  Prevention. 2007;Available from:
  \url{https://www.cdc.gov/std/chlamydia/chlamydiascreening-males.pdf}.

\bibitem[{Gift et~al.(2008)Gift, Thomas L. and Gaydos, Charlotte A. and Kent,
  Charlotte K. and Marrazzo, Jeanne M. and Rietmeijer, Cornelis A. and
  Schillinger, Julia A. and Dunne, Eileen F.}]{gift2008program}
Gift TL, Gaydos CA, Kent CK, Marrazzo JM, Rietmeijer CA, Schillinger JA, et~al.
\newblock The program cost and cost-effectiveness of screening men for
  chlamydia to prevent pelvic inflammatory disease in women.
\newblock Sex Transm Dis. 2008 Nov;35(11 Suppl):S66--75.

\bibitem[{Gopalappa et~al.(2013)Gopalappa, Chaitra and Huang, Y A and Gift,
  Thomas L and {Owusu-Edusei}, Kwame and Taylor, Melanie and Gales,
  Vincent}]{gopalappa2013cost}
Gopalappa C, Huang YA, Gift TL, {Owusu-Edusei} K, Taylor M, Gales V.
\newblock Cost-effectiveness of screening men in {{Maricopa County}} jails for
  chlamydia and gonorrhea to avert infections in women.
\newblock Sex Transm Dis. 2013 Oct;40(10):776--783.

\bibitem[{Kissinger et~al.(2019)Kissinger, Patricia and Schmidt, Norine and
  Gomes, G{\'e}rard and {Craig-Kuhn}, Megan Clare and Scott, Glenis and Watson,
  Shannon and Lederer, Alyssa}]{KissingerA71}
Kissinger P, Schmidt N, Gomes G, {Craig-Kuhn} MC, Scott G, Watson S, et~al.
\newblock O14.2 {{Can}} community {{Chlamydia}} trachomatis screening of young
  heterosexual men help identify infected networks?
\newblock Sex Transm Infect. 2019;95(Suppl 1):A71--A71.
\newblock Available from: \url{https://sti.bmj.com/content/95/Suppl_1/A71.1}.
  DOI:10.1136/sextrans-2019-sti.183.

\bibitem[{Green et~al.(2014)Green, Jakevia and Schmidt, Norine and Latimer,
  Jennifer and Johnson, Taylor and Aktaruzzaman, Upama and Flanigan, Emily and
  Olugbade, Yewande and Bangel, Steffani and Clum, Gretchen and Madkour, Aubrey
  and Johnson, Carolyn and Kissinger, Patricia}]{greenWP58}
Green J, Schmidt N, Latimer J, Johnson T, Aktaruzzaman U, Flanigan E, et~al.
\newblock {{WP}} 58 {{The}} influence of partnership type and characteristics
  on condom use among young {{African American}} women.
\newblock Sex Transm Dis. 2014 Jun;41(Suppl 1):S111--S111.
\newblock Available from:
  \url{https://www.cdc.gov/stdconference/2014/2014-std-prevention-conference-abstracts.pdf}.
  DOI:10.1097/01.olq.0000451608.27700.9f.

\bibitem[{Azizi et~al.(2019)Azizi, Asma and Qu, Zhuolin and Lewis, Bryan and
  Hyman, James M.}]{azizi2018generating}
Azizi A, Qu Z, Lewis B, Hyman JM.
\newblock Generating a heterosexual bipartite network embedded in social
  network.
\newblock Submitted. 2019;.

\bibitem[{{Network Dynamics and Simulation and Science
  Laboratory}(2008)}]{NDSSL2008synthetic}
{Network Dynamics and Simulation and Science Laboratory}.
\newblock Synthetic data products for societal infrastructures and
  protopopulations: {{Data}} set 2.0.
\newblock {Virginia Polytechnic Institute and State University}; 2008.
  NDSSL-TR-07-003.
\newblock Available from:
  \url{http://ndssl.vbi.vt.edu/Publications/ndssl-tr-07-003.pdf}.

\bibitem[{Farley et~al.(2003)Farley, Thomas A and Cohen, Deborah A and Elkins,
  Whitney}]{farley2003asymptomatic}
Farley TA, Cohen DA, Elkins W.
\newblock Asymptomatic sexually transmitted diseases: the case for screening.
\newblock Prev Med. 2003 Apr;36(4):502--509.

\bibitem[{{Centers for Disease Control and
  Prevention}(????{\natexlab{a}})}]{CDCept}
{Centers for Disease Control and Prevention}.
\newblock Sexually {{Transmitted Diseases}} ({{STDs}}), {{Expedited Partner
  Therapy}}; [updated 2019 Aug 26; cited 2020 Jan 4].
\newblock Available from: \url{https://www.cdc.gov/std/ept/default.htm}.

\bibitem[{{Centers for Disease Control and
  Prevention}(????{\natexlab{b}})}]{CDCtreat}
{Centers for Disease Control and Prevention}.
\newblock Chlamydia {{Treatment}} and {{Care}}; [updated 2016 Dec 9; cited 2020
  Jan 4].
\newblock Available from:
  \url{https://www.cdc.gov/std/chlamydia/treatment.htm}.

\bibitem[{St.~Lawrence et~al.(2002)St. Lawrence, Janet S. and Monta{\~n}o,
  Daniel E. and Kasprzyk, Danuta and Phillips, William R. and Armstrong, Keira
  and Leichliter, Jami S.}]{st.lawrence2002STD}
St~Lawrence JS, Monta{\~n}o DE, Kasprzyk D, Phillips WR, Armstrong K,
  Leichliter JS.
\newblock {{STD}} screening, testing, case reporting, and clinical and partner
  notification practices: a national survey of {{US}} physicians.
\newblock Am J Public Health. 2002 Nov;92(11):1784--1788.

\bibitem[{Torrone et~al.(2014)Torrone, Elizabeth and Papp, John and Weinstock,
  Hillard}]{torrone2014prevalence}
Torrone E, Papp J, Weinstock H.
\newblock Prevalence of {{Chlamydia}} trachomatis genital infection among
  persons aged 14-39 years--{{United States}}, 2007-2012.
\newblock MMWR Morb Mortal Wkly Rep. 2014 Sep;63(38):834--838.

\bibitem[{Molano et~al.(2005)Molano, M{\'o}nica and Meijer, Chris J. L. M. and
  Weiderpass, Elisabete and Arslan, Annie and Posso, Hector and Franceschi,
  Silvia and Ronderos, Margarita and Mu{\~n}oz, Nubia and {van den Brule},
  Adriaan J. C.}]{molano2005natural}
Molano M, Meijer CJLM, Weiderpass E, Arslan A, Posso H, Franceschi S, et~al.
\newblock The natural course of {{Chlamydia}} trachomatis infection in
  asymptomatic {{Colombian}} women: a 5-year follow-up study.
\newblock J Infect Dis. 2005 Mar;191(6):907--916.

\bibitem[{Chitnis et~al.(2008)Chitnis, Nakul and Hyman, James M and Cushing,
  Jim M}]{chitnis2008determining}
Chitnis N, Hyman JM, Cushing JM.
\newblock Determining important parameters in the spread of malaria through the
  sensitivity analysis of a mathematical model.
\newblock Bull Math Biol. 2008 Feb;70(5):1272--1296.

\bibitem[{Kretzschmar et~al.(1996)Kretzschmar, M. and {van Duynhoven}, Y. T. H.
  P. and Severijnen, A. J.}]{kretzschmar1996modeling}
Kretzschmar M, {van Duynhoven} YTHP, Severijnen AJ.
\newblock Modeling prevention strategies for gonorrhea and chlamydia using
  stochastic network simulations.
\newblock Am J Epidemiol. 1996 Aug;144(3):306--317.

\bibitem[{Turner et~al.(2006)Turner, KatherineME and Adams, ElisabethJ and Gay,
  Nigel and Ghani, AzraC and Mercer, Catherine and Edmunds, W
  John}]{turner2006developing}
Turner K, Adams E, Gay N, Ghani A, Mercer C, Edmunds WJ.
\newblock Developing a realistic sexual network model of chlamydia transmission
  in {{Britain}}.
\newblock Theor Biol Med Model. 2006 Jan;3(1):3.

\bibitem[{Martin et~al.(1992)Martin, David H and Mroczkowski, Tomasz F and
  Dalu, ZA and McCarty, James and Jones, Robert B and Hopkins, Scott J and
  Johnson, Raymond B and GROUP*, AZITHROMYCIN FOR CHLAMYDIAL INFECTIONS
  STUDY}]{martin1992controlled}
Martin DH, Mroczkowski TF, Dalu Z, McCarty J, Jones RB, Hopkins SJ, et~al.
\newblock A controlled trial of a single dose of azithromycin for the treatment
  of {{Chlamydial}} urethritis and cervicitis.
\newblock N Engl J Med. 1992 Sep;327(13):921--925.

\bibitem[{Trussell(2002)Trussell, James}]{trussell2007choosing}
Trussell J.
\newblock Choosing a contraceptive: efficacy, safety, and personal
  considerations.
\newblock In: Hatcher R, Trussell J, Nelson A, Cates W, Stewart F, Kowal D,
  editors. Contraceptive technology. 19th ed. New York: Ardent Media, Inc;
  2002. p. 19--47.

\bibitem[{{Centers for Disease Control and Prevention}(????)}]{CDCdetail}
{Centers for Disease Control and Prevention}.
\newblock Chlamydia, {{Detailed Fact Sheet}}; [updated 2016 Oct 4; cited 2020
  Jan 4].
\newblock Available from:
  \url{https://www.cdc.gov/std/chlamydia/stdfact-chlamydia-detailed.htm}.

\bibitem[{Wiehe et~al.(2011)Wiehe, Sarah E. and Rosenman, Marc B. and Wang,
  Jane and Katz, Barry P. and Fortenberry, J. Dennis}]{wiehe2011chlamydia}
Wiehe SE, Rosenman MB, Wang J, Katz BP, Fortenberry JD.
\newblock Chlamydia screening among young women: individual- and provider-level
  differences in testing.
\newblock Pediatrics. 2011 Feb;127(2):e336--e344.

\bibitem[{Hoover et~al.(2014)Hoover, Karen W. and Leichliter, Jami S. and
  Torrone, Elizabeth A. and Loosier, Penny S. and Gift, Thomas L. and Tao,
  Guoyu}]{hoover2014chlamydia}
Hoover KW, Leichliter JS, Torrone EA, Loosier PS, Gift TL, Tao G.
\newblock Chlamydia screening among females aged 15 - 21 years -- multiple data
  sources, {{United States}}, 1999 - 2010.
\newblock Morb Mortal Wkly Rep. 2014 Sep;63(Suppl 2):80--88.

\bibitem[{Hogben et~al.(2005)Hogben, Matthew and McCree, Donna H. and Golden,
  Matthew R.}]{hogben2005patient}
Hogben M, McCree DH, Golden MR.
\newblock Patient-delivered partner therapy for sexually transmitted diseases
  as practiced by {{U}}.{{S}}. physicians.
\newblock Sex Transm Dis. 2005 Feb;32(2):101--105.

\bibitem[{Golden et~al.(2005)Golden, Matthew R. and Whittington, William L.H.
  and Handsfield, H. Hunter and Hughes, James P. and Stamm, Walter E. and
  Hogben, Matthew and Clark, Agnes and Malinski, Cheryl and Helmers, Jennifer
  R.L. and Thomas, Katherine K. and Holmes, King K.}]{golden2005effect}
Golden MR, Whittington WLH, Handsfield HH, Hughes JP, Stamm WE, Hogben M,
  et~al.
\newblock Effect of expedited treatment of sex partners on recurrent or
  persistent gonorrhea or chlamydial infection.
\newblock N Engl J Med. 2005 Feb;352(7):676--685.

\bibitem[{Schillinger et~al.(2003)Schillinger, Julia A. and Kissinger, Patricia
  and Calvet, Helene and Whittington, William L. H. and Ransom, Ray L. and
  Sternberg, Maya R. and Berman, Stuart M. and Kent, Charlotte K. and Martin,
  David H. and Oh, M. Kim and Handsfield, H. Hunter and Bolan, Gail and
  Markowitz, Lauri E. and Fortenberry, J. Dennis}]{schillinger2003patient}
Schillinger JA, Kissinger P, Calvet H, Whittington WLH, Ransom RL, Sternberg
  MR, et~al.
\newblock Patient-delivered partner treatment with azithromycin to prevent
  repeated {{Chlamydia}} trachomatis infection among women: a randomized,
  controlled trial.
\newblock Sex Transm Dis. 2003 Jan;30(1):49--56.

\bibitem[{Xu et~al.(2011)Xu, Fujie and Stoner, Bradley P. and Taylor, Stephanie
  N. and Mena, Leandro and Tian, Lin H. and Papp, John and Hutchins, Kathleen
  and Martin, David H. and Markowitz, Lauri E.}]{xu2011use}
Xu F, Stoner BP, Taylor SN, Mena L, Tian LH, Papp J, et~al.
\newblock Use of home-obtained vaginal swabs to facilitate rescreening for
  {{Chlamydia}} trachomatis infections: two randomized controlled trials.
\newblock Obstet Gynecol. 2011 Aug;118(2):231--239.

\bibitem[{{RelayHealth}(????)}]{Cttest}
{RelayHealth}.
\newblock {Gonorrhea and Chlamydia Tests}; [updated 2014 Jan 01; cited 2020 Feb
  2].
\newblock Available from:
  \url{https://www.summitmedicalgroup.com/library/adult_health/aha_gonorrhea_and_chlamydia_tests}.

\bibitem[{Batteiger et~al.(2010)Batteiger, Byron E. and Xu, Fujie and Johnson,
  Robert E. and Rekart, Michael L.}]{batteiger2010Protective}
Batteiger BE, Xu F, Johnson RE, Rekart ML.
\newblock Protective immunity to {{Chlamydia}} trachomatis genital infection:
  evidence from human studies.
\newblock J Infect Dis. 2010 Jun;201(S2):178--189.

\end{thebibliography}

\end{document}